\documentclass[journal,final,dvipdfmx]{IEEEtran}

\usepackage{cite}
\usepackage{graphicx}
\usepackage[cmex10]{amsmath}
\usepackage{url}
\usepackage{color}

\usepackage{amssymb}
\usepackage{amsfonts}
\usepackage{multirow}
\usepackage{bm}

\makeatletter
\newcommand{\textarrow}[2][1]
  { \settowidth{\@tempdima}{#2}
    \stackrel{#2}
             {\makebox[#1\@tempdima][l]{\rightarrowfill}}
  }
\makeatother

\def\Hline{\noalign{\hrule height 0.4mm}}

\newcommand{\PreserveBackslash}[1]{\let\temp=\\#1\let\\=\temp}

\newcommand{\revise}[1]{\textcolor{black}{#1}}
\title{Singing Voice Separation and Vocal F0 Estimation based on Mutual Combination of 
  Robust Principal Component Analysis and Subharmonic Summation}

\author{Yukara~Ikemiya,~\IEEEmembership{Student Member,~IEEE,}
        Katsutoshi~Itoyama,~\IEEEmembership{Member,~IEEE,}
        and Kazuyoshi~Yoshii,~\IEEEmembership{Member,~IEEE}% <-this % stops a space
\thanks{The authors are with the Department
of Intelligence Science and Technology, Graduate School of Informatics,
Kyoto University, Kyoto, Japan
(e-mail:\{ikemiya, itoyama, yoshii\}@kuis.kyoto-u.ac.jp).}% <-this % stops a space
}

\begin{document}
\fussy

\maketitle

\begin{abstract}
This paper presents a new method of singing voice analysis 
 that performs mutually-dependent singing voice separation 
 and vocal fundamental frequency (F0) estimation.
Vocal F0 estimation is considered to become easier
 if singing voices can be separated from a music audio signal,
 and vocal F0 contours are useful for singing voice separation.
This calls for an approach that improves the performance of each of these tasks
 by using the results of the other. 
The proposed method first performs robust principal component analysis (RPCA) 
 for roughly extracting singing voices from a target music audio signal. 
The F0 contour of the main melody is then estimated from the separated singing voices 
 by finding the optimal temporal path over an F0 saliency spectrogram.
Finally, the singing voices are separated again more accurately
 by combining a conventional time-frequency mask given by RPCA 
 with another mask that passes only the harmonic structures of the estimated F0s. 
Experimental results showed that the proposed method
 significantly improved the performances of both singing voice separation and vocal F0 estimation.
The proposed method also outperformed all the other methods of singing voice separation
 submitted to an international music analysis competition called MIREX 2014. 
\end{abstract}


% \fi
\IEEEpeerreviewmaketitle

\section{Introduction}
\fussy

\IEEEPARstart{S}{inging} voice analysis is important 
 for active music listening interfaces\cite{goto:2007}
 that enable a user to customize the contents of existing music recordings
 in ways not limited to frequency equalization and tempo adjustment.
Since singing voices tend to form main melodies
 and strongly affect the moods of musical pieces,
 several methods have been proposed for editing
 the three major kinds of acoustic characteristics of singing voices:
 fundamental frequencies (F0s), timbres, and volumes.
A system of speech analysis and synthesis called TANDEM-STRAIGHT~\cite{straight},
 for example, decomposes human voices
 into F0s, spectral envelopes
 (timbres), and non-periodic components.
High-quality F0- and/or timbre-changed singing voices
 can then be resynthesized by manipulating F0s and spectral envelopes.
Ohishi {\it et al.}~\cite{ohishi:2014}
 represents F0 or volume dynamics of singing voices
 by using a probabilistic model
 and transfers those dynamics to other singing voices.
Note that these methods deal only with isolated singing voices. 
Fujihara and Goto~\cite{fujihara:2011}
 model the spectral envelopes of singing voices in polyphonic audio signals
 to directly modify the vocal timbres
 without affecting accompaniment parts.

To develop a system that enables a user to edit   
 the acoustic characteristics of singing voices included in a polyphonic audio signal,
 we need to accurately perform {\it both} singing voice separation and vocal F0 estimation.
The performance of each task could be improved
 by using the results of the other
 because there is a complementary relationship between them.
If singing voices were extracted from a polyphonic audio signal, 
 it would be easy to estimate a vocal F0 contour from them.
Vocal F0 contours are useful for improving singing voice separation.
In most studies, however, only the {\it one-way} dependency 
 between the two tasks has been considered.
Singing voice separation has often been used 
 as preprocessing for vocal F0 estimation, and vice versa.

In this paper 
 we propose a novel singing voice analysis method 
 that performs singing voice separation and vocal F0 estimation
 in an interdependent manner.
The core component of the proposed method 
 is preliminary singing voice separation based 
 on robust principal component analysis (RPCA)~\cite{huang:2012}.
Given the amplitude spectrogram (matrix) of a music signal,
 RPCA decomposes it into the sum of a low-rank matrix and a sparse matrix.
Since accompaniments
 such as drums and rhythm guitars
 tend to play similar phrases repeatedly,
 the resulting spectrogram generally has a low-rank structure.
Since singing voices vary significantly and continuously over time
 and the power of singing voices concentrates on harmonic partials,
 on the other hand,
 the resulting spectrogram has a not low-rank but sparse structure.
Although RPCA is considered to be one of the most prominent ways of singing voice separation,
 non-repetitive instrument sounds 
 are inevitably assigned to a sparse spectrogram.
To filter out such non-vocal sounds,
 we estimate the F0 contour of singing voices
 from the sparse spectrogram
 based on a saliency-based F0 estimation method called subharmonic summation (SHS) \cite{shs}
 and extract only a series of harmonic structures 
 corresponding to the estimated F0s.
Here we propose a novel F0 saliency spectrogram in the time-frequency domain
 by leveraging the results of RPCA.
This can avoid
 the negative effect of accompaniment sounds in vocal F0 estimation.

\begin{figure}[t]
  \centering
  \includegraphics[width=0.9\columnwidth]{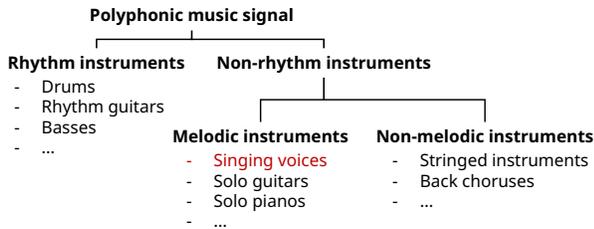}
  \vspace{-5pt}
  \caption{Typical instrumental composition of popular music.
  \label{popular_inst}}
  \vspace{-15pt}
\end{figure}

Our method is similar in spirit to a recent method of singing voice separation
 that combines rhythm-based and pitch-based methods of singing voice separation \cite{rafii:2014}.
It first estimates
 two types of {\it soft} time-frequency masks passing only singing voices
 by using a singing voice separation method called REPET-SIM \cite{repet_sim}
 and a vocal F0 estimation method (originally proposed for multiple-F0 estimation \cite{duan:ieee:2010}).
Those soft masks are then integrated into a unified mask in a weighted manner.
On the other hand,
our method is deeply linked to 
human perception of a main melody 
in polyphonic music~\cite{palmer:1987,friberg:2009}.
Fig. \ref{popular_inst} shows an instrumental composition of popular music.
It is thought that humans easily recognize the sounds of rhythm instruments 
 such as drums and rhythm guitars~\cite{palmer:1987} 
 and that in the residual sounds of non-rhythm instruments, 
 spectral components that have predominant harmonic structures are identified 
 as main melodies~\cite{friberg:2009}.
The proposed method first separates the sounds of rhythm instruments 
 by using a time-frequency (TF) mask estimated by RPCA.
Main melodies are extracted as singing voices
 from the residual sounds by using another mask that passes only predominant harmonic structures.
\revise{
Although
 the main melodies do not always correspond to singing voices,
 we do not deal with vocal activity detection (VAD) in this paper
 because many promising VAD methods~\cite{ramona:2008,fujihara:2011:journal,lehner:2015}
 can be applied as pre- or post-processing of our method.
}

The rest of this paper is organized as follows. 
Section \ref{related_work} introduces related works. 
Section \ref{proposed_method} explains the proposed method.
Section \ref{evaluation} describes the evaluation experiments
 and the MIREX 2014 singing-voice-separation task results.
Section \ref{system_parameters} describes the experiments 
determining robust parameters for the proposed method.
Section \ref{conclusion} concludes this paper.

\section{Related Work}
\label{related_work}

This section introduces related works
on vocal F0 estimation and singing voice separation.
It also reviews some studies on the combination of those two tasks.

\subsection{Vocal F0 Estimation}

A typical approach to vocal F0 estimation
 is to identify F0s that have predominant harmonic structures
 by using an F0 saliency spectrogram
 that represents how likely the F0 is to exist in each time-frequency bin.
A core of this approach is how to estimate
a saliency spectrogram~\cite{goto:2004,rao:2010,dressler:2011,arora:2012,salamon:2012}.
Goto~\cite{goto:2004} proposed 
 a statistical multiple-F0 analyzer called {\it PreFEst}
 that approximates an observed spectrum
 as a superimposition of harmonic structures.
Each harmonic structure is represented as a Gaussian mixture model (GMM)
 and the mixing weights of GMMs corresponding to different F0s 
 can be regarded as
 a saliency spectrum.
Rao {\it et al.} \cite{rao:2010} 
 tracked multiple candidates of vocal F0s
 including the F0s of locally predominant non-vocal sounds
 and then identified vocal F0s 
 by focusing on the temporal instability of vocal components.
Dressler \cite{dressler:2011} 
 attempted to reduce the number of possible overtones
 by identifying which overtones are derived from a vocal harmonic structure.
Salamon {\it et al.} \cite{salamon:2012}
proposed a heuristics-based method called {\it MELODIA} that
 focuses on the characteristics of vocal F0 contours.
The contours of F0 candidates are obtained by using a saliency spectrogram
based on subharmonic summation.
This method achieved the state-of-the-art results in vocal F0 estimation.

\subsection{Singing Voice Separation}

A typical approach to singing voice separation
 is to make a TF mask that separates a target music spectrogram
 into a vocal spectrogram and an accompaniment spectrogram.
There are two types of TF masks: soft masks and binary masks.
An ideal binary mask assigns 1 to a TF
 unit if the power of singing voices in the unit 
 is larger than that of the other concurrent sounds, and 0 otherwise.
Although vocal and accompaniment sounds overlap with various ratios at many TF units, 
 excellent separation can be achieved using binary masking.
This is related to a phenomenon called auditory masking: 
 a louder sound tends to mask a weaker sound within a particular frequency band \cite{wang:2005}.

Nonnegative matrix factorization (NMF) has often been used
 for separating a polyphonic spectrogram into nonnegative components
 and clustering those components 
 into vocal components and accompaniment components \cite{chanrungutai:2008,zhu:2013,yang:2014}.
Another approach is to exploit the temporal and spectral continuity of accompaniment sounds 
 and the sparsity of singing voices in the TF domain
 \cite{tachibana:2014,fitzgerald:2010,jeong:2014}.
Tachibana {\it et al.} \cite{tachibana:2014}, for example,
 proposed harmonic/percussive source separation (HPSS)
 based on the isotropic natures of harmonic and percussive sounds.
Both components were estimated jointly
 via maximum a posteriori (MAP) estimation.
Fitzgerald {\it et al.} \cite{fitzgerald:2010}
 proposed an HPSS method applying different median filters 
 to polyphonic spectra along the time and frequency directions.
Jeong {\it et al.} \cite{jeong:2014} statistically modeled
 the continuities of accompaniment sounds and the sparsity of singing voices.
Yen {\it et al.} \cite{yen:2014} separated vocal, harmonic, and percussive components
 by clustering frequency modulation features in an unsupervised manner.
Huang {\it et al.} \cite{huang:2014} have recently used a deep recurrent neural network
 for supervised singing voice separation.

Some state-of-the-art methods of singing voice separation 
 focus on the repeating characteristics of accompaniment sounds \cite{rafii:2013,repet_sim,huang:2012}.
Accompaniment sounds are often played by musical instruments 
 that repeat similar phrases throughout the music, such as drums and rhythm guitars.
To identify repetitive patterns in a polyphonic audio signal, 
 Rafii {\it et al.} \cite{rafii:2013} took the median of repeated spectral segments 
 detected by an autocorrelation method, 
 and improved the separation by using a similarity matrix \cite{repet_sim}.
Huang {\it et al.} \cite{huang:2012} used RPCA to 
 identify repetitive structures of accompaniment sounds.
Liutkus {\it et al.} \cite{liutkus:2014} proposed kernel additive modeling
% that subsumes many conventional methods 
 that \revise{combines} many conventional methods 
 and accounts for various features like continuity, smoothness, and stability over time or frequency.
These methods tend to work robustly in several situations or genres
 because they make few assumptions about the target signal.
\revise{
Driedger {\it et al.} \cite{driedger:2015}
 proposed a cascading method
 that first decomposes a music spectrogram
 into harmonic, percussive, and residual spectrograms,
 each of which is further decomposed
 into partial components of singing voices and those of accompaniment sounds 
 by using conventional methods \cite{virtanen:2008,huang:2014}.
Finally, the estimated components are reassembled
 to form singing voices and accompaniment sounds.
}

\subsection{One-way or Mutual Combination}

Since singing voice separation and vocal F0 estimation have complementary relationships,
 the performance of each task can be improved by using the results of the other.
Some vocal F0 estimation methods
 use singing voice separation techniques as preprocessing 
 for reducing the negative effect of accompaniment sounds 
 in polyphonic music \cite{tachibana:2014,hsu:2010,yeh:2012,rafii:2013}. 
This approach results in comparatively better performance
 when the volume of singing voices is relatively low \cite{salamon:2014}.
Some methods of singing voice separation
 use vocal F0 estimation techniques because the energy of a singing voice 
 is concentrated on an F0 and its harmonic partials \cite{li:2007,virtanen:2008,fujihara:2010}.
Virtanen {\it et al.} \cite{virtanen:2008} proposed a method that first separates harmonic components using a predominant F0 contour.
The residual components are then modeled by NMF and accompaniment sounds are extracted.
Singing voices and accompaniment sounds are separated by using the learned parameters again.

Some methods perform both vocal F0 estimation and singing voice separation.
Hsu {\it et al.} \cite{Hsu:2012}
 proposed a tandem algorithm that iterates these two tasks.
Durrieu {\it et al.} \cite{durrieu:2011} 
 used source-filter NMF for directly modeling 
 the F0s and timbres of singing voices and accompaniment sounds.
% and separated each source by estimating model parameters.
Rafii {\it et al.} \cite{rafii:2014}
 proposed a framework that combines 
 repetition-based source separation
 with F0-based source separation.
A unified TF mask for singing voice separation is
 obtained by combining the TF masks estimated by the two types of source separation in a weighted manner.
\revise{
Caba{\~{n}}as-Molero {\it et al.} \cite{cabanas:2011}
 proposed a method
 that roughly separates singing voices from stereo recordings by focusing on the spatial diversity
 (called {\it center extraction})
 and then estimates a vocal F0 contour for the separated voices.
The separation of singing voices is further improved by using the F0 contour.
}

% =================
%  Proposed Method
% =================
\section{Proposed Method}
\label{proposed_method}

\begin{figure}[t]
  \centering
  \includegraphics[width=0.9\columnwidth]{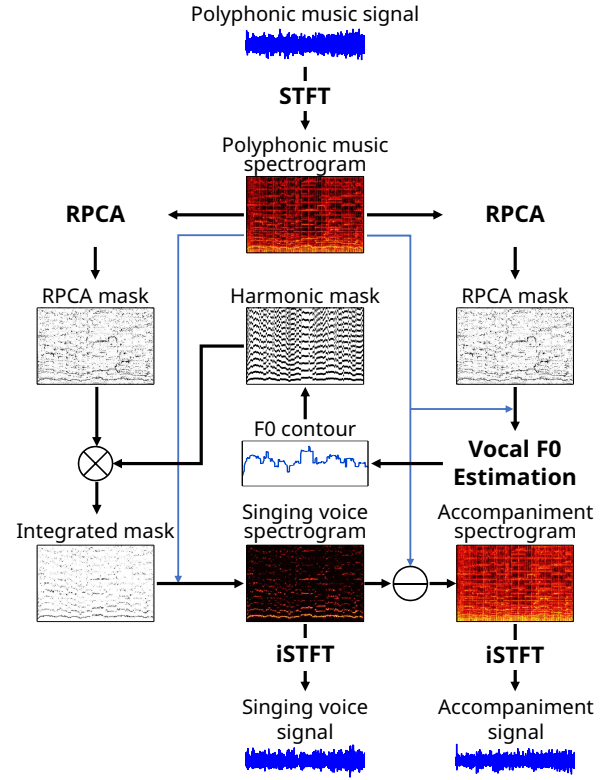}
  \vspace{-5pt}
  \caption{Overview of the proposed method.
    First an RPCA mask that separates low-rank components in a polyphonic spectrogram is computed.
    From this mask and the original spectrogram, a vocal F0 contour is estimated.
    The RPCA mask and the harmonic mask calculated from the F0 contour are combined by multiplication,
    and finally the singing voice and the accompaniment sounds are separated using the integrated mask.
  \label{overview}}
  \vspace{-10pt}
\end{figure}

% The proposed method mutually executes 
The proposed method \revise{jointly} executes 
 singing voice separation and vocal F0 estimation (Fig.~\ref{overview}).
Our method uses robust principal component analysis (RPCA) 
 to estimate a mask (called an RPCA mask)
 that separates a target music spectrogram 
 into low-rank components and sparse components.
The vocal F0 contour is then estimated 
 from the separated sparse components 
 via Viterbi search on an F0 saliency spectrogram,
 resulting in another mask (called a harmonic mask)
 that separates harmonic components of the estimated F0 contour.
These masks are integrated via element-wise multiplication,
and finally singing voices and accompaniment sounds are 
obtained by separating the music spectrogram
according to the integrated mask.
The proposed method
 can work well for complicated music audio signals.
Even if the volume of singing voices is relatively low
 and music audio signals contain various kinds of musical instruments,
 the harmonic structures (F0s) of singing voices can be discovered
 by calculating an F0 saliency spectrogram from an RPCA mask.

\subsection{Singing Voice Separation}

Vocal and accompaniment sounds are separated 
by combining TF masks based on RPCA and vocal F0s.

\subsubsection{Calculating an RPCA Mask}
\label{rpca-based method}

A singing voice separation method based on RPCA \cite{huang:2012}
assumes that
 accompaniment and vocal components tend to have low-rank and sparse structures, respectively,
 in the TF domain.
Since spectra of harmonic instruments ({\it e.g.,} pianos and guitars)
 are consistent for each F0
 and the F0s are basically discretized at a semitone level,
 harmonic spectra having the same shape
 appear repeatedly in the same musical piece.
Spectra of non-harmonic instruments ({\it e.g.}, drums)
 also tend to appear repeatedly.
\revise{
Vocal spectra, in contrast, rarely have the same shape 
 because the vocal timbres and F0s
 vary continuously and significantly over time.
}%

RPCA decomposes an input matrix $\bm{X}$ 
 into the sum of a low-rank matrix $\bm{X}_L$ and a sparse matrix $\bm{X}_S$
 by solving the following convex optimization problem:
\begin{align}
  \mbox{minimize} \;\; &\|\bm{X}_L\|_* + \hat{\lambda}\|\bm{X}_S\|_1 \;\; 
  (\mbox{subject to}\;\; \bm{X}_L + \bm{X}_S = \bm{X}), \nonumber\\
  &\hat{\lambda} = \frac{\lambda}{\sqrt{\max(T, F)}},
\end{align}
where $\bm{X}$, $\bm{X}_L$, and $\bm{X}_S \in \mathbb{R}^{T\times F}$,
 $\|\cdot\|_*$ and $\|\cdot\|_1$ 
% represent the nuclear norm and the L1-norm, respectively.
 represent the \revise{nuclear norm (also known as the trace norm)} and the L1-norm, respectively.
$\lambda$ is a positive parameter that controls the balance 
 between the low-rankness of $\bm{X}_L$ and the sparsity of $\bm{X}_S$.
To find optimal $\bm{X}_L$ and $\bm{X}_S$, 
 we use an efficient inexact version 
 of the augmented Lagrange multiplier (ALM) algorithm \cite{lin:2009}.

When $\bm{X}$ is the amplitude spectrogram 
 given by the short-time Fourier transform (STFT) of a target music audio signal 
 ($T$ is the number of frames and $F$ is the number of frequency bins),
 the spectral components having repetitive structures are assigned to $\bm{X}_L$
 and the other varying components are assigned to $\bm{X}_S$.
Let $t$ and $f$ be a time frame and a frequency bin, respectively ($1 \le t \le T$ and $1 \le f \le F$).
\revise{
We obtain a TF soft mask $\bm{M}_{\mbox{\tiny RPCA}}^{(\mathrm{s})} \in \mathbb{R}^{T\times F}$
by using Wiener filtering:
\begin{align}
  M_{\mbox{\tiny RPCA}}^{(\mathrm{s})}(t, f) = \frac{|X_S(t, f)|}{|X_S(t, f)| + |X_L(t, f)|}.
\end{align}
A TF binary mask $\bm{M}_{\mbox{\tiny RPCA}}^{(\mathrm{b})} \in \mathbb{R}^{T\times F}$ is also obtained 
by comparing $\bm{X}_L$ with $\bm{X}_S$ in an element-wise manner as follows:
%% We obtain a TF binary mask $\bm{M}_{\mbox{\tiny RPCA}} \in \mathbb{R}^{T\times F}$
%%  by comparing $\bm{X}_L$ with $\bm{X}_S$ in an element-wise manner as follows:
\begin{align}
  M_{\mbox{\tiny RPCA}}^{(\mathrm{b})}(t, f) = \left \{
  \begin{array}{l}
    1 \;\;\;\; \mbox{if} \;\; |X_S(t, f)| > \gamma |X_L(t, f)| \\
    0 \;\;\;\; \mbox{otherwise}
  \end{array}
  \right..
\end{align}
The gain $\gamma$ adjusts the energy between the low-rank and sparse matrices.
In this paper the gain parameter is set to 1.0,
 which was reported to achieve good separation performance \cite{huang:2012}.
Note that $\bm{M}_{\mbox{\tiny RPCA}}^{(\mathrm{b})}$ is used only for 
estimating a vocal F0 contour in Section \ref{vocal_f0_estimation}.
}

% Using the mask, 
Using $\bm{M}_{\mbox{\tiny RPCA}}^{(\mathrm{s})}$ or $\bm{M}_{\mbox{\tiny RPCA}}^{(\mathrm{b})}$,
 the vocal spectrogram $\bm{X}_{\mbox{\tiny VOCAL}}^{(*)} \in \mathbb{R}^{T\times F}$
 is roughly estimated as follows:
\begin{align}
  \bm{X}_{\mbox{\tiny VOCAL}}^{(*)} = \bm{M}_{\mbox{\tiny RPCA}}^{(*)} \odot \bm{X},
\end{align}
 where $\odot$ indicates the element-wise product.
\revise{If the value of $\lambda$ for singing voice separation
 is different from that for F0 estimation}, 
 we execute two versions of RPCA 
 with different values of $\lambda$ (Fig. \ref{overview}).
If we were to use the same value of $\lambda$ for both processes, 
 RPCA would be executed only once.
In section \ref{system_parameters} 
 we discuss the optimal values of $\lambda$ in detail.

\subsubsection{Calculating a Harmonic Mask}
\label{harmonic_mask}

Using a vocal F0 contour $\hat{\bm{Y}} = \{\hat{y_1},\hat{y_2},\cdots,\hat{y_T}\}$
 (see details in Section \ref{vocal_f0_estimation}),
 we make a harmonic mask $\bm{M}_{\mbox{\tiny H}} \in \mathbb{R}^{T\times F}$.
\revise{
Assuming that the energy of vocal spectra is localized 
 on the harmonic partials of vocal F0s,
we defined $\bm{M}_{\mbox{\tiny H}} \in \mathbb{R}^{T\times F}$ as:
\begin{align}
  \bm{M}_{\mbox{\tiny H}}(t, f) 
  = \left \{
  \begin{array}{l}
    \mathrm{w}(n; W) \;\; \mbox{if} \;\; 
    \begin{array}{l}
      % W_{\mathrm{l}}^h \le n \le W_{\mathrm{u}}^h, \\
      0 < f - w_{\mathrm{u}}^n \le W, \\
      w_{\mathrm{l}}^n = \mathrm{f}(n h_{\hat{y_t}} - \frac{w}{2}), \\
      w_{\mathrm{u}}^n = \mathrm{f}(n h_{\hat{y_t}} + \frac{w}{2}), \\
      W = w_{\mathrm{l}}^n - w_{\mathrm{u}}^n + 1,
    \end{array}\\
    0 \;\;\;\;\;\;\;\;\;\;\;\;\; \mbox{otherwise},
  \end{array}
  \right.
\end{align}
where $\mathrm{w}(n; W)$ denotes the $n$-th value of a window function of length $W$,
$\mathrm{f}(h)$ denotes the index of the nearest time frame corresponding to a frequency $h$ [Hz],
$n$ is the index of a harmonic partial,
$w$ is a frequency width [Hz] for extracting the energy around the partial,
$h_{\hat{y_t}}$ is the estimated vocal F0 [Hz] of frame $t$.
We chose the Tukey window whose a shape parameter is set to 0.5 as a window function.
}
%% Assuming that the energy of vocal spectra is localized 
%%  on the harmonic partials of vocal F0s,
%% $\bm{M}_{\mbox{\tiny H}} \in \mathbb{R}^{T\times F}$ can be obtained by
%% \begin{align}
%%   M_{\mbox{\tiny H}}(t, f) 
%%   = \left \{
%%   \begin{array}{l}
%%     1 \;\; \mbox{if} \;\; n h_{y_t^*} - \frac{w}{2} < h_f < n h_{y_t^*} + \frac{w}{2}, \\
%%     0 \;\; \mbox{otherwise},
%%   \end{array}
%%   \right.
%% \end{align}
%% where $n$ is the index of a harmonic partial,
%%  $w$ is a frequency width [Hz] for extracting the energy around the partial,
%%  $h_{y_t^*}$ is the estimated vocal F0 [Hz] of frame $t$,
%%  and $h_f$ is the frequency [Hz] corresponding to frequency bin $f$.

\subsubsection{Integrating the Two Masks for Singing Voice Separation}

\revise{
Given the RPCA mask (soft) $\bm{M}_{\mbox{\tiny RPCA}}^{(\mathrm{s})}$ 
 and the harmonic mask $\bm{M}_{\mbox{\tiny H}}$,
 we define an integrated soft mask $\bm{M}_{\mbox{\tiny RPCA+H}}^{(\mathrm{s})}$ as follows:
\begin{align}
  \bm{M}_{\mbox{\tiny RPCA+H}}^{(\mathrm{s})} 
  = \bm{M}_{\mbox{\tiny RPCA}}^{(\mathrm{s})} \odot \bm{M}_{\mbox{\tiny H}}.
\end{align}
%% define binary mask
Furthermore, 
an integrated binary mask $\bm{M}_{\mbox{\tiny RPCA+H}}^{(\mathrm{b})}$
is also defined as:
\begin{align}
  \bm{M}_{\mbox{\tiny RPCA+H}}^{(\mathrm{b})}(t, f) = \left \{
  \begin{array}{l}
    1 \;\;\;\; \mbox{if} \;\; \bm{M}_{\mbox{\tiny RPCA+H}}^{(\mathrm{s})}(t, f) > 0.5 \\
    0 \;\;\;\; \mbox{otherwise}.
  \end{array}
  \right..
\end{align}
}%
Although the integrated masks have fewer spectral units 
 assigned to singing voices than the RPCA mask and the harmonic mask do, 
 they provide better separation quality
 (see the comparative results reported in Section \ref{system_parameters}).

Using the integrated masks $\bm{M}_{\mbox{\tiny RPCA+H}}^{(*)}$,
 the vocal and accompaniment spectrograms $\hat{\bm{X}}_{\mbox{\tiny VOCAL}}^{(*)}$ 
 and $\hat{\bm{X}}_{\mbox{\tiny ACCOM}}^{(*)}$ are given by
\begin{align}
  \hat{\bm{X}}_{\mbox{\tiny VOCAL}}^{(*)} &= \bm{M}_{\mbox{\tiny RPCA+H}}^{(*)} \odot \bm{X}, \nonumber \\
  \hat{\bm{X}}_{\mbox{\tiny ACCOM}}^{(*)} &= \bm{X} - \hat{\bm{X}}_{\mbox{\tiny VOCAL}}^{(*)}.
\end{align}
Finally, time signals (waveforms) of singing voices and accompaniment sounds
 are resynthesized by computing the inverse STFT with the phases of the original music spectrogram.

%
% Figure
%
\begin{figure}[t]
  \centering
  \includegraphics[width=0.9\columnwidth]{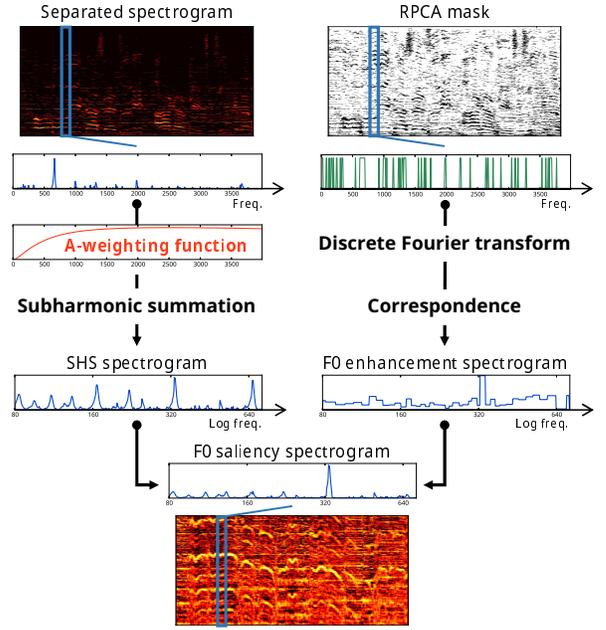}
  \vspace{-5pt}
  \caption{An F0-saliency spectrogram is obtained by integrating
    an SHS spectrogram derived from a separated vocal spectrogram
    with an F0 enhancement spectrogram
    derived from an RPCA mask.
  \label{salience_comp}}
  \vspace{-10pt}
\end{figure}

\subsection{Vocal F0 Estimation}
\label{vocal_f0_estimation}

We propose a new method that estimates a vocal F0 contour $\hat{\bm{Y}} = \{\hat{y_1},\cdots,\hat{y_T}\}$
 from the vocal spectrogram $\bm{X}_{\mbox{\tiny VOCAL}}^{(\mathrm{b})}$
 by using the binary mask $\bm{M}_{\mbox{\tiny RPCA}}^{(\mathrm{b})}$.
A robust F0-saliency spectrogram is obtained by
 using both $\bm{X}_{\mbox{\tiny VOCAL}}^{(\mathrm{b})}$ and $\bm{M}_{\mbox{\tiny RPCA}}^{(\mathrm{b})}$
 and a vocal F0 contour is estimated 
 by finding an optimal path in the saliency spectrogram
 with the Viterbi search algorithm.

\subsubsection{Calculating a Log-frequency Spectrogram}
\label{spectral_transform}

We convert the vocal spectrogram $\bm{X}_{\mbox{\tiny VOCAL}}^{(\mathrm{b})} \in \mathbb{R}^{T \times F}$
 to the log-frequency spectrogram $\bm{X}'_{\mbox{\tiny VOCAL}} \in \mathbb{R}^{T \times C}$
 by using spline interpolation on the dB scale.
A frequency $h_f$ [Hz] is translated to 
 the index of a log-frequency bin $c \ (1 \le c \le C)$ as follows:
\begin{align}
  c = \left\lfloor \frac{1200 \log_{2} \frac{h_f}{h_{\mathrm{low}}}}{p} + 1 \right\rfloor,
\end{align}
where $h_{\mathrm{low}}$ is 
 a predefined lowest frequency [Hz] and $p$ a frequency resolution [cents] per bin.
The frequency $h_{\mathrm{low}}$ must be sufficiently low 
 to include the low end of a singing voice spectrum ({\it i.e.}, 30 Hz).

To take into account the non-linearity of human auditory perception,
 we multiply the A-weighting function $R_A(f)$ 
 to the vocal spectrogram $\bm{X}_{\mbox{\tiny VOCAL}}^{(\mathrm{b})}$ in advance.
$R_A(f)$ is given by
\begin{align}
  R_A(f) =& \frac{12200^2 h_f^4}{(h_f^2 + 20.6^2)(h_f^2 + 12200^2)} \nonumber \\
  & \times \frac{1}{\sqrt{(h_f^2 + 107.7^2)(h_f^2 + 737.9^2)}}.
\end{align}
This function is a rough approximation of the inverse of the 40-phon equal-loudness curve\footnote{http://replaygain.hydrogenaud.io\/proposal\/equal\_loudness.html}
and is used for amplifying the frequency bands that we are perceptually sensitive to,
 and attenuating the frequency bands that we are less sensitive to \cite{salamon:2012}.

\subsubsection{Calculating an F0-Saliency Spectrogram}
\label{salience_f0est}

Fig. \ref{salience_comp} shows the procedure of 
calculating an F0-Saliency spectrogram.
We calculate a subharmonic summation (SHS) spectrogram $\bm{S}_{\mbox{\tiny SHS}} \in \mathbb{R}^{T \times C}$
 from the tentative vocal spectrogram $\bm{X}'_{\mbox{\tiny VOCAL}} \in \mathbb{R}^{T \times C}$
 in the log-frequency domain.
SHS \cite{shs} is the most basic and light-weight algorithm
 that underlies many vocal F0 estimation methods \cite{cao:2007,salamon:2012}.
$\bm{S}_{\mbox{\tiny SHS}}$ is given by
\begin{align}
  S_{\mbox{\tiny SHS}}(t, c) 
  = \sum_{n=1}^{N} \beta_n \bm{X}'_{\mbox{\tiny VOCAL}} \left(t, c + \left\lfloor\frac{1200\log_2n}{p}\right\rfloor \right),
\end{align}
where $c$ is the index of a log-frequency bin ($1 \le c \le C$),
$N$ is the number of harmonic partials considered, 
 and $\beta_n$ is a decay factor ($0.86^{n-1}$ in this paper).

We then calculate an F0 enhancement spectrogram $\bm{S}_{\mbox{\tiny RPCA}} \in \mathbb{R}^{T \times C}$ 
 from the RPCA mask $\bm{M}_{\mbox{\tiny RPCA}}$.
To improve the performance of vocal F0 estimation,
 we propose to focus on the regularity (periodicity) of harmonic partials over the linear frequency axis.
The RPCA binary mask $\bm{M}_{\mbox{\tiny RPCA}}$ can be used for reducing half or double pitch errors
 because the harmonic structure of the singing voice strongly appears in it.

We first take the discrete Fourier transform (DFT) of each time frame of the binary mask as follows:
\begin{align}
 F(t, k) = \left|\sum_{f=0}^{F-1}M_{\mbox{\tiny RPCA}}^{(\mathrm{b})}(t, f) e^{-i\frac{2\pi kf}{F}}\right|.
\end{align}
\revise{%
This idea is similar to the cepstral analysis
 that extracts the periodicity of harmonic partials from log-power spectra. 
We do not need to compute the log of the RPCA binary mask
 because $\bm{M}_{\mbox{\tiny RPCA}} \in \{0,1\}^{T \times F}$. 
}%
The F0 enhancement spectrogram $\bm{S}_{\mbox{\tiny RPCA}}$ 
 is obtained by picking the value corresponding to a frequency index $c$:
\begin{align}
  S_{\mbox{\tiny RPCA}}(t,c) &= F\left(t, \left\lfloor\frac{h_{\mathrm{top}}}{h_c}\right\rfloor\right),
\end{align}
where $h_c$ is the frequency [Hz] 
 corresponding to log-frequency bin $c$ and
 $h_{\mathrm{top}}$ is the highest frequency [Hz] considered (Nyquist frequency).

Finally, the reliable F0-saliency spectrogram $\bm{S} \in \mathbb{R}^{T \times C}$
 is given by integrating $\bm{S}_{\mbox{\tiny SHS}}$
 and $\bm{S}_{\mbox{\tiny RPCA}}$ as follows:
\begin{align}
  S(t,c) = S_{\mbox{\tiny SHS}}(t,c) S_{\mbox{\tiny RPCA}}(t,c)^\alpha,
\end{align}
where $\alpha$ is a weighting factor for adjusting the balance 
 between $\bm{S}_{\mbox{\tiny SHS}}$ and $\bm{S}_{\mbox{\tiny RPCA}}$.
When $\alpha$ is 0, $\bm{S}_{\mbox{\tiny RPCA}}$ is ignored,
 resulting in the standard SHS method.
While each bin of $\bm{S}_{\mbox{\tiny SHS}}$ reflects the total volume of harmonic partials,
 each bin of $\bm{S}_{\mbox{\tiny RPCA}}$ reflects the number of harmonic partials.

\subsubsection{Executing Viterbi Search}
\label{viterbi_search}

Given the F0-saliency spectrogram $\bm{S}$, 
 we estimate the optimal F0 contour $\hat{\bm{Y}} = \{\hat{y_1},\cdots,\hat{y_T}\}$
 by solving the following problem:
\begin{align}
    \hat{\bm{Y}} = \underset{y_1,...,y_T}{\mbox{argmax}}
        \sum_{t=1}^{T - 1} \left\{ \log \frac{\bm{S}(t, y_t)}{\sum_{c=c_l}^{c_h} \bm{S}(t, c)} + \log G(y_t, y_{t+1})
        \right\},
\end{align}
where $c_l$ and $c_h$ are the lowest and highest log-frequency bins of an F0 search range. 
$G(y_t, y_{t+1})$ is the transition cost function
 from the current F0 $y_t$ to the next F0 $y_{t+1}$.
$G(y_t, y_{t+1})$ is defined as
\begin{align}
  G(y_t, y_{t+1}) = \frac{1}{2b}\exp \left(-\frac{|c_{y_t} - c_{y_{t+1}}|}{b}\right).
\end{align}
where $b = \sqrt{\frac{150^2}{2}}$ and $c_y$ indicates the log-frequency [cents] 
 corresponding to log-frequency bin $c$.
This function is equivalent to the Laplace distribution whose standard deviation is 150 [cents].
Note that the shifting interval of time frames is 10 [ms].
This optimization problem can be efficiently solved using the Viterbi search algorithm.

%
% Table
%

\begin{table*}[t]
  \centering
  \caption{Datasets and parameters}
  \label{tab:setting}
  \begin{tabular}{c|ccccccccc}
  \Hline
     & Number of clips & Length of clips & Sampling rate & Window size & Hopsize & $N$ & $\lambda$ & $w$ & $\alpha$ \\ \hline
    MIR-1K & 110 & 20--110 sec & 16 kHz & 2048 & 160 & 10 & 0.8 & 50 & 0.6 \\
    MedleyDB    & 45 & 17--514 sec & 44.1 kHz & 4096 & 441 & 20 & 0.8 & 70 & 0.6 \\
    RWC-MDB-2001    & 100 & 125--365 sec & 44.1 kHz & 4096 & 441 & 20 & 0.8 & 70 & 0.6 \\
  \Hline
  \end{tabular}
\end{table*}

\begin{table}[t]
  \centering
  \caption{Song clips in {\it MedleyDB} used for evaluation.
  \label{medleydb_used_clips}}
  \begin{tabular}{rp{0.53\columnwidth}}
    \Hline
    Artists                  & Songs                 \\
    \hline
    A Classic Education     & Night Owl            \\
    Aimee Norwich           & Child                \\
    Alexander Ross          & Velvet Curtain       \\
    Auctioneer              & Our Future Faces     \\
    Ava Luna                & Waterduct            \\
    Big Troubles            & Phantom              \\
    Brandon Webster         & Dont Hear A Thing, Yes Sir I Can Fly    \\
    Clara Berry And Wooldog & Air Traffic, Boys, Stella, Waltz For My Victims          \\
    Creepoid                & Old Tree             \\
    Dreamers Of The Ghetto  & Heavy Love           \\
    Faces On Film           & Waiting For Ga       \\
    Family Band             & Again                \\
    Helado Negro            & Mitad Del Mundo      \\
    Hezekiah Jones          & Borrowed Heart       \\
    Hop Along               & Sister Cities        \\
    Invisible Familiars     & Disturbing Wildlife  \\
    Liz Nelson              & Coldwar, Rainfall              \\
    Matthew Entwistle       & Dont You Ever        \\
    Meaxic                  & Take A Step, You Listen            \\
    Music Delta             & 80s Rock, Beatles, Britpop, Country1, Country2, Disco, Gospel, Grunge, Hendrix, Punk, Reggae, Rock, Rockabilly  \\
    Night Panther           & Fire                 \\
    Port St Willow          & Stay Even            \\
    Secret Mountains        & High Horse           \\
    Steven Clark            & Bounty               \\
    Strand Of Oaks          & Spacestation         \\
    Sweet Lights            & You Let Me Down      \\
    The Scarlet Brand       & Les Fleurs Du Mal    \\
    \Hline
  \end{tabular}

\end{table}

\section{Experimental Evaluation}
\label{evaluation}

This section reports
 experiments conducted for evaluating singing voice separation and vocal F0 estimation.
The results of the {\it Singing Voice Separation} task of MIREX 2014, 
which is a world-wide competition between algorithms for music analysis,
are also shown.

\subsection{Singing Voice Separation}
\label{experiment_sep}

Singing voice separation using different binary masks was evaluated to 
verify the effectiveness of the proposed method.

%
% figure
%
\begin{figure*}[t]
  \centering
  {\large \textbf{MIR-1K}}\\
  \vspace{5pt}
  \begin{tabular}{c}
    \centering
    \begin{minipage}{.33\textwidth}
      \centering
      Singing voice\\
      \includegraphics[width=0.99\columnwidth]{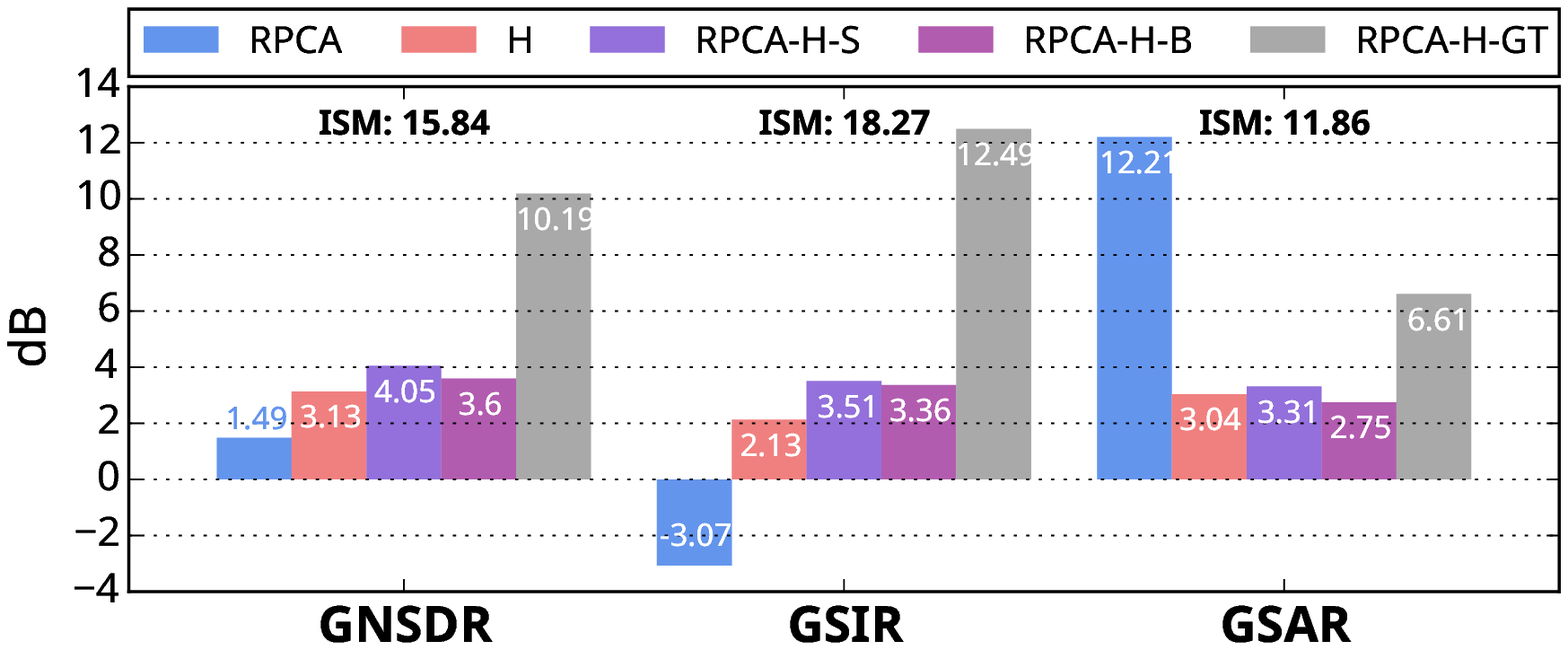}\\
      \vspace{-5pt}
      Accompaniment\\
      \vspace{-10pt}
      \includegraphics[width=0.99\columnwidth]{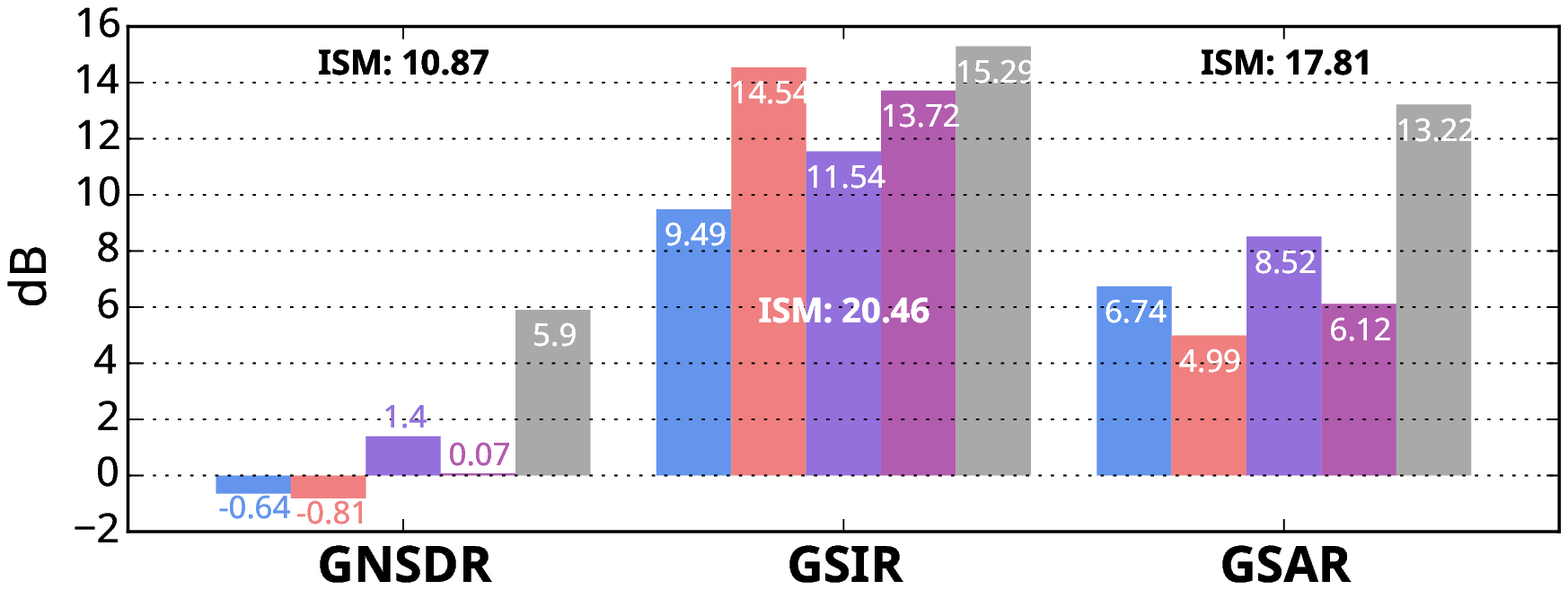}\\
      \text{(a) $-5$ dB SNR}
    \end{minipage}
    \begin{minipage}{.33\textwidth}
      \centering
      Singing voice\\
      \includegraphics[width=0.99\columnwidth]{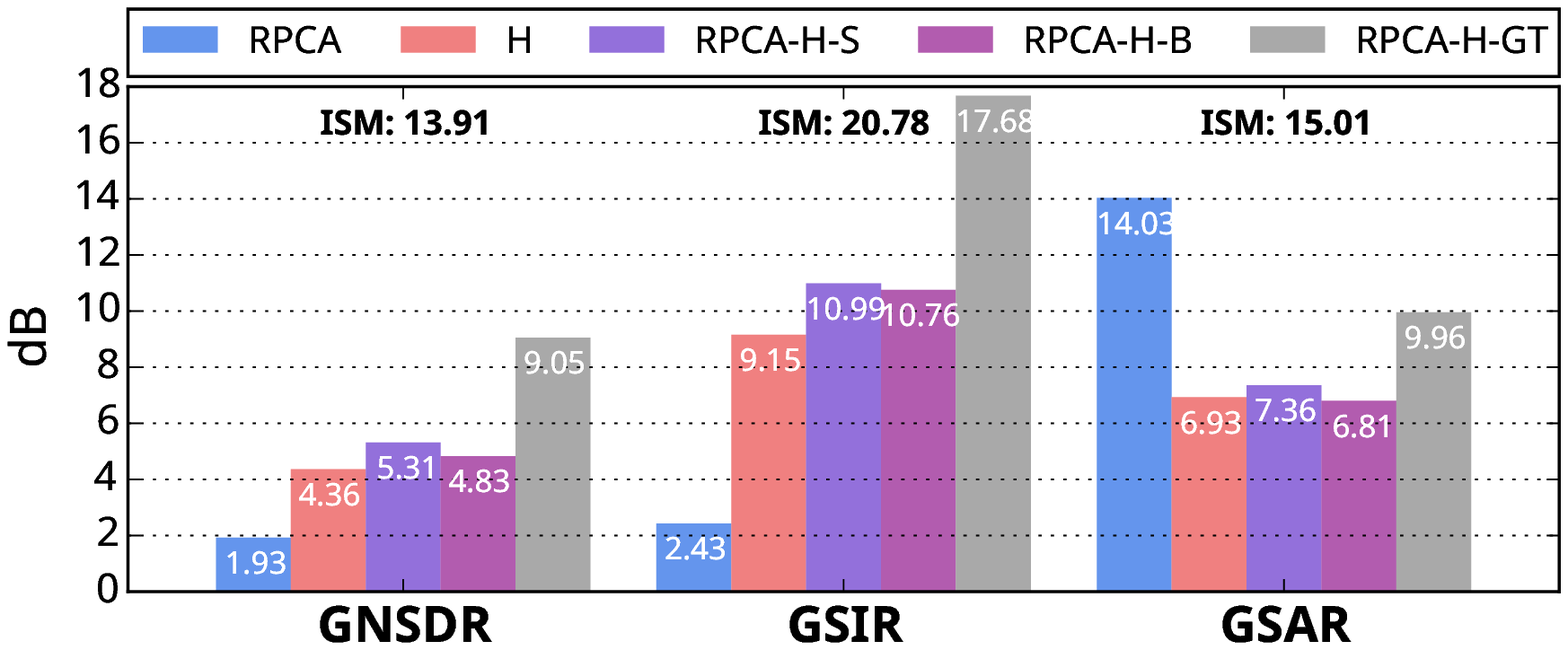}\\
      \vspace{-5pt}
      Accompaniment\\
      \vspace{-10pt}
      \includegraphics[width=0.99\columnwidth]{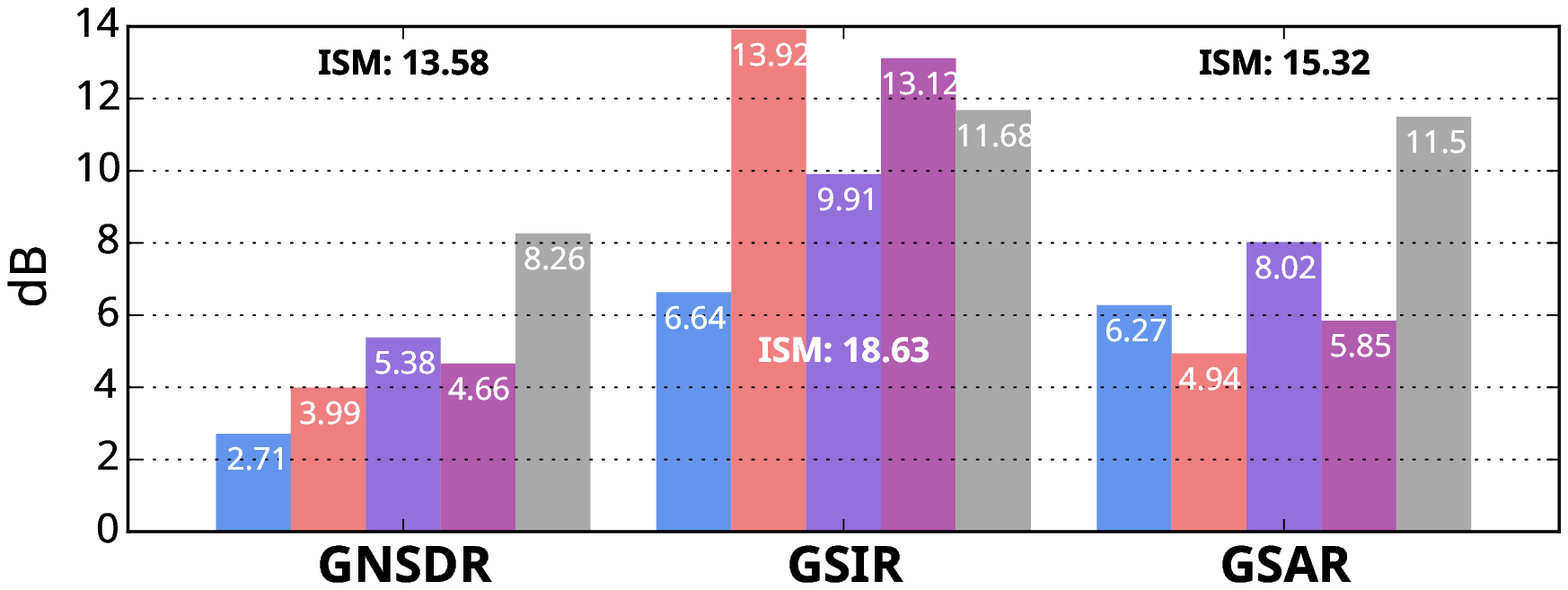}\\
      \text{(b) $0$ dB SNR}
    \end{minipage}
    \begin{minipage}{.33\textwidth}
      \centering
      Singing voice\\
      \includegraphics[width=0.99\columnwidth]{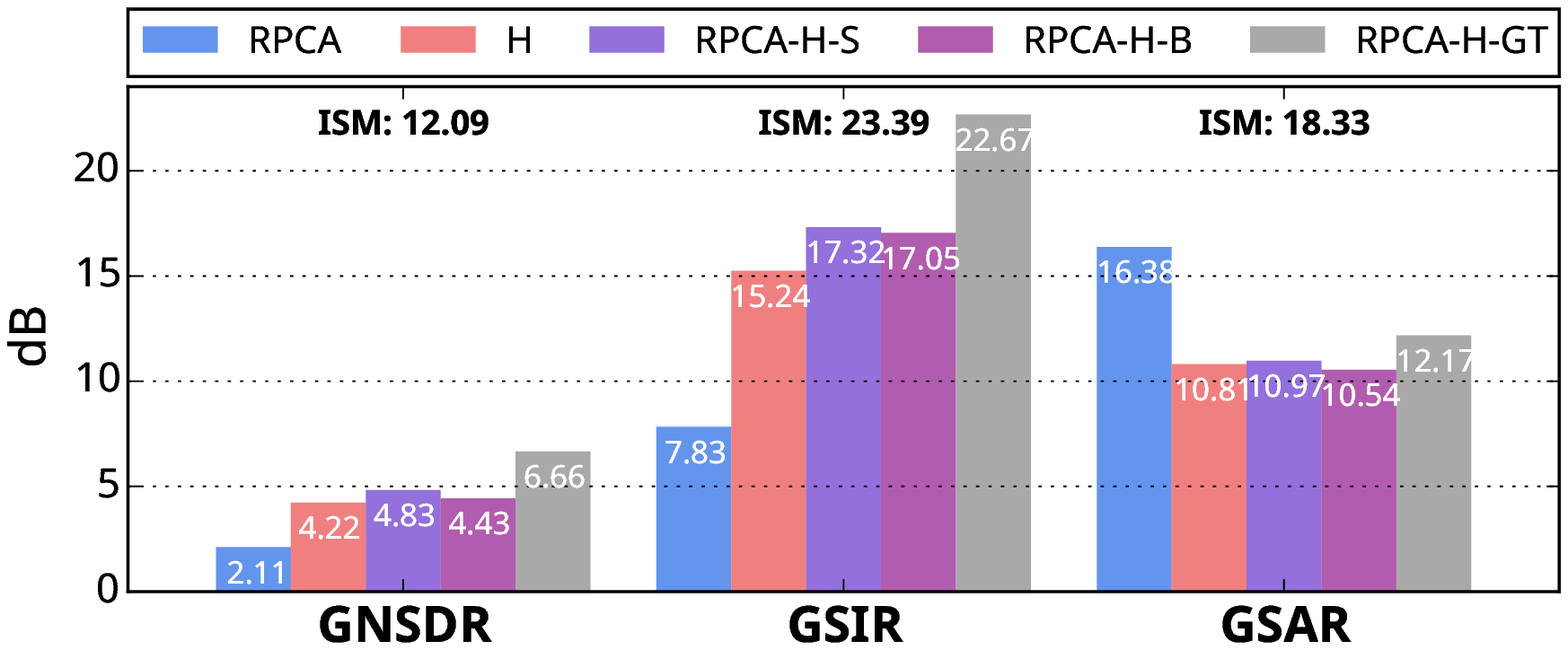}\\
      \vspace{-5pt}
      Accompaniment\\
      \vspace{-10pt}
      \includegraphics[width=0.99\columnwidth]{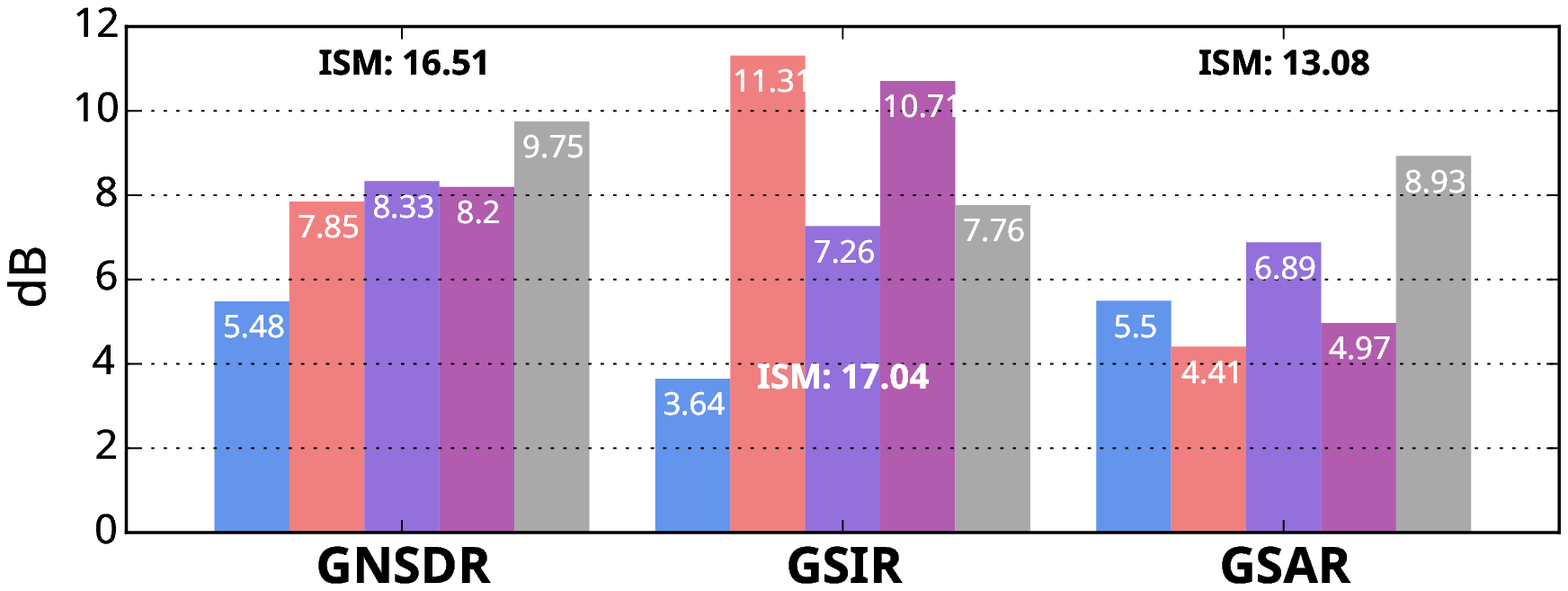}\\
      \text{(c) $5$ dB SNR}
    \end{minipage}
  \end{tabular}\\
  \vspace{8pt}
  {\large \textbf{MedleyDB}}\\
  \vspace{5pt}
  \begin{tabular}{c}
    \begin{minipage}{.33\textwidth}
      \centering
      Singing voice\\
      \includegraphics[width=0.99\columnwidth]{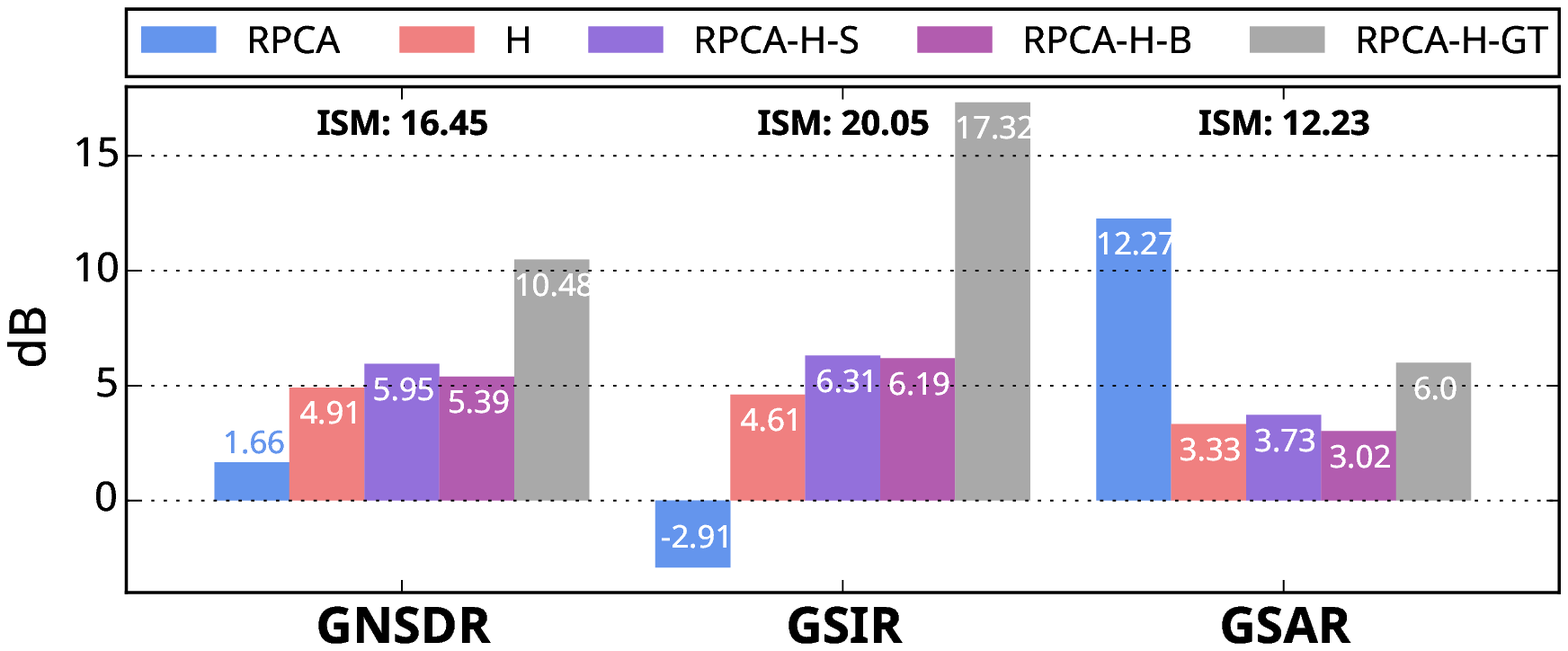}\\
      \vspace{-5pt}
      Accompaniment\\
      \vspace{-10pt}
      \includegraphics[width=0.99\columnwidth]{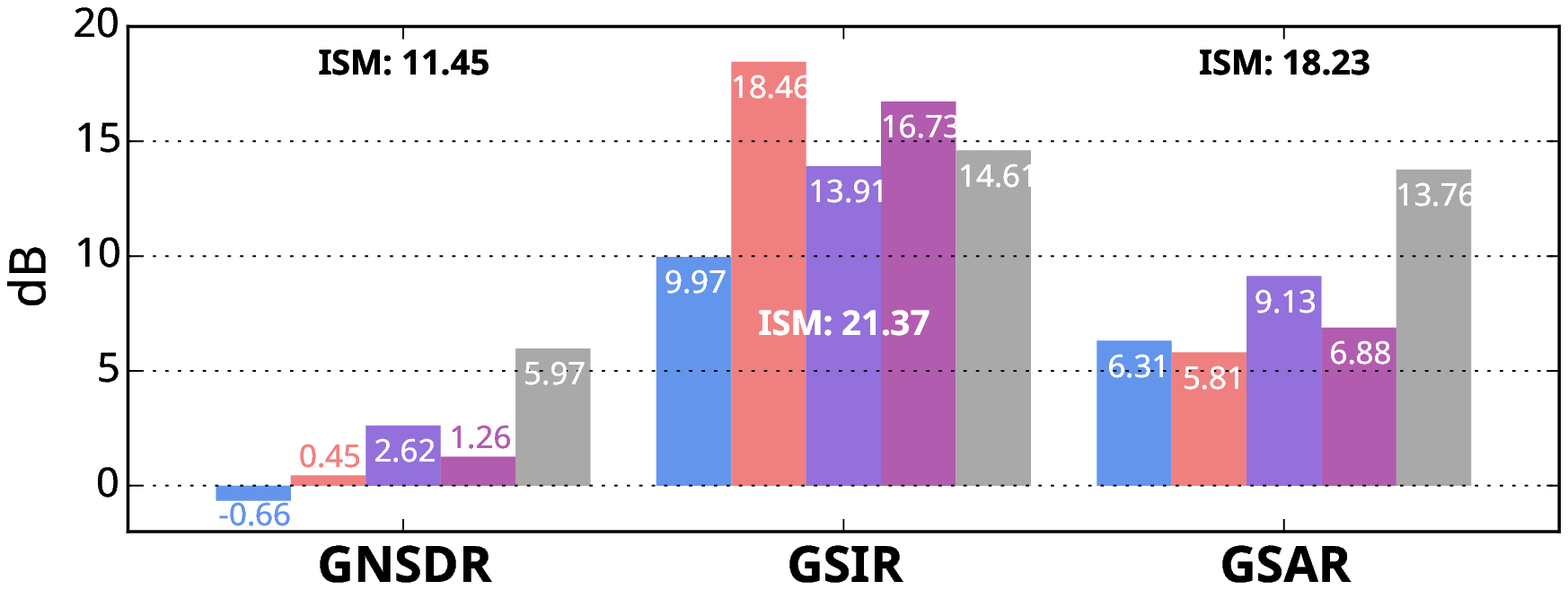}\\
      \text{(a) $-5$ dB SNR}
    \end{minipage}
    \begin{minipage}{.33\textwidth}
      \centering
      Singing voice\\
      \includegraphics[width=0.99\columnwidth]{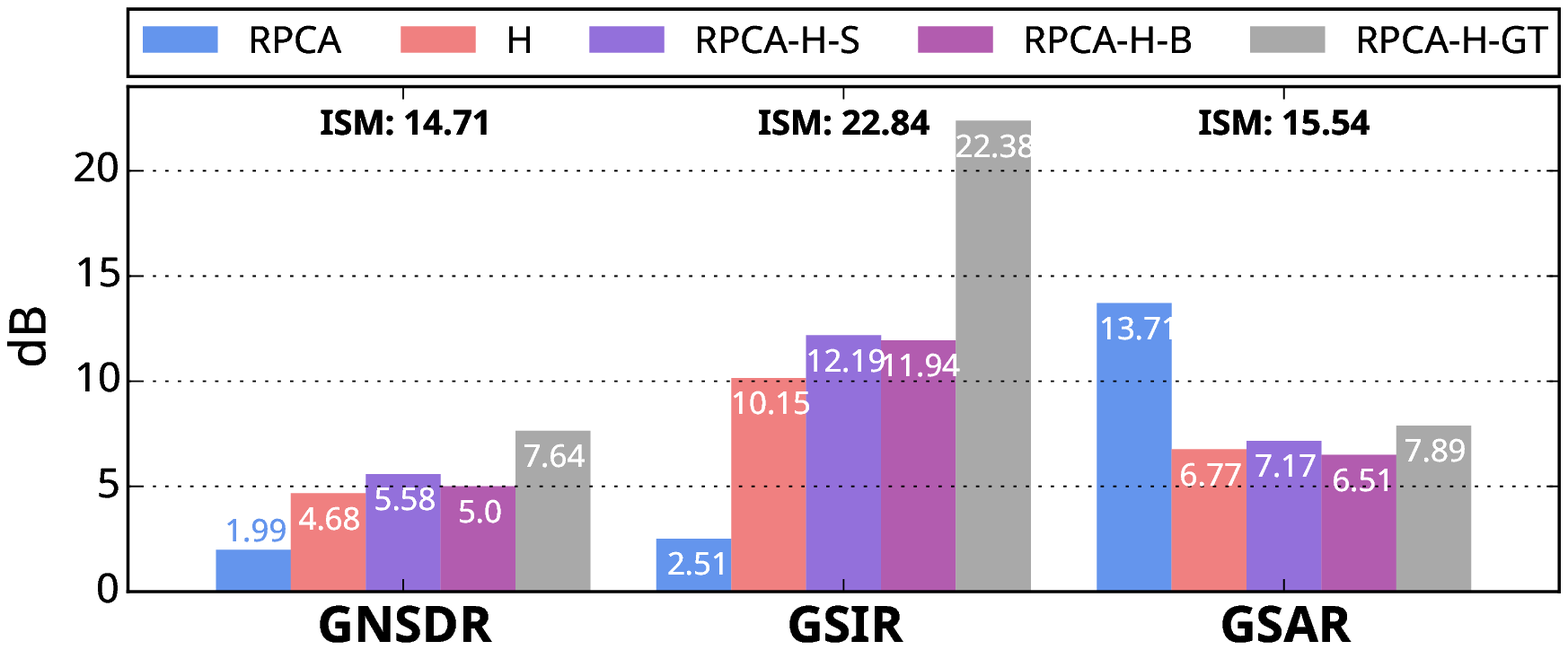}\\
      \vspace{-5pt}
      Accompaniment\\
      \vspace{-10pt}
      \includegraphics[width=0.99\columnwidth]{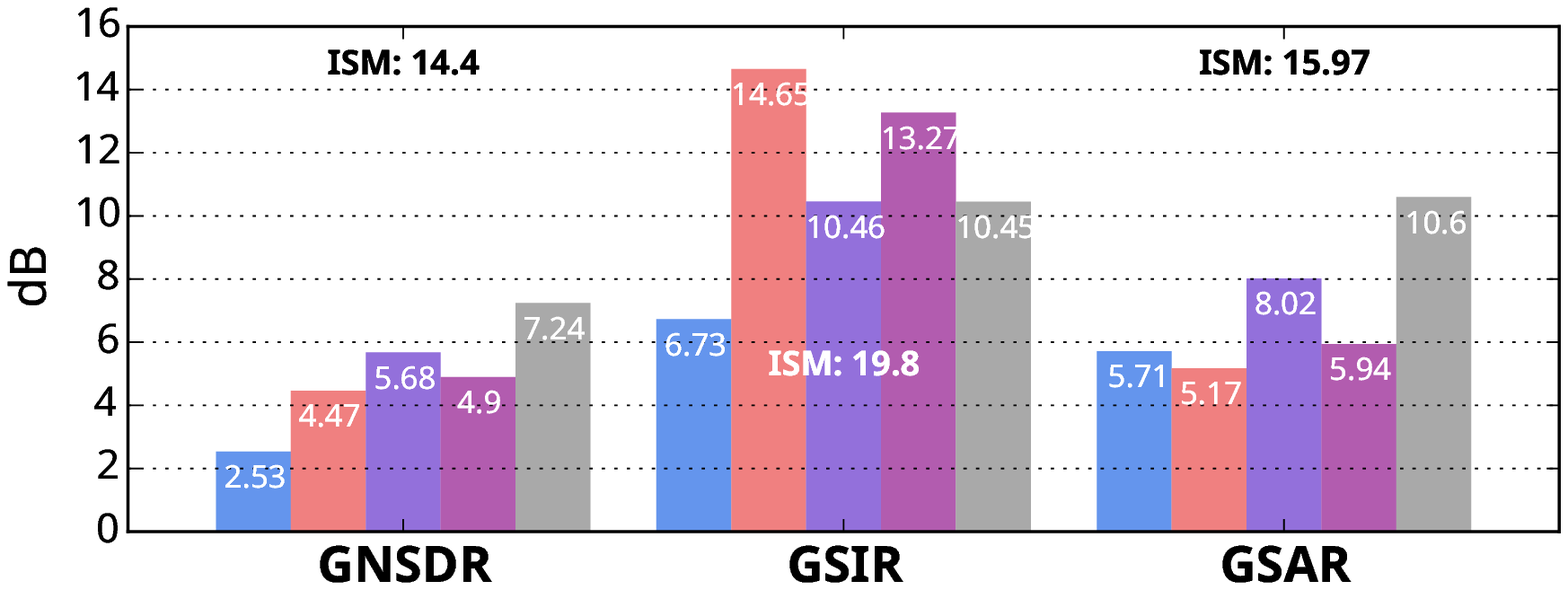}\\
      \text{(b) $0$ dB SNR}
    \end{minipage}
    \begin{minipage}{.33\textwidth}
      \centering
      Singing voice\\
      \includegraphics[width=0.99\columnwidth]{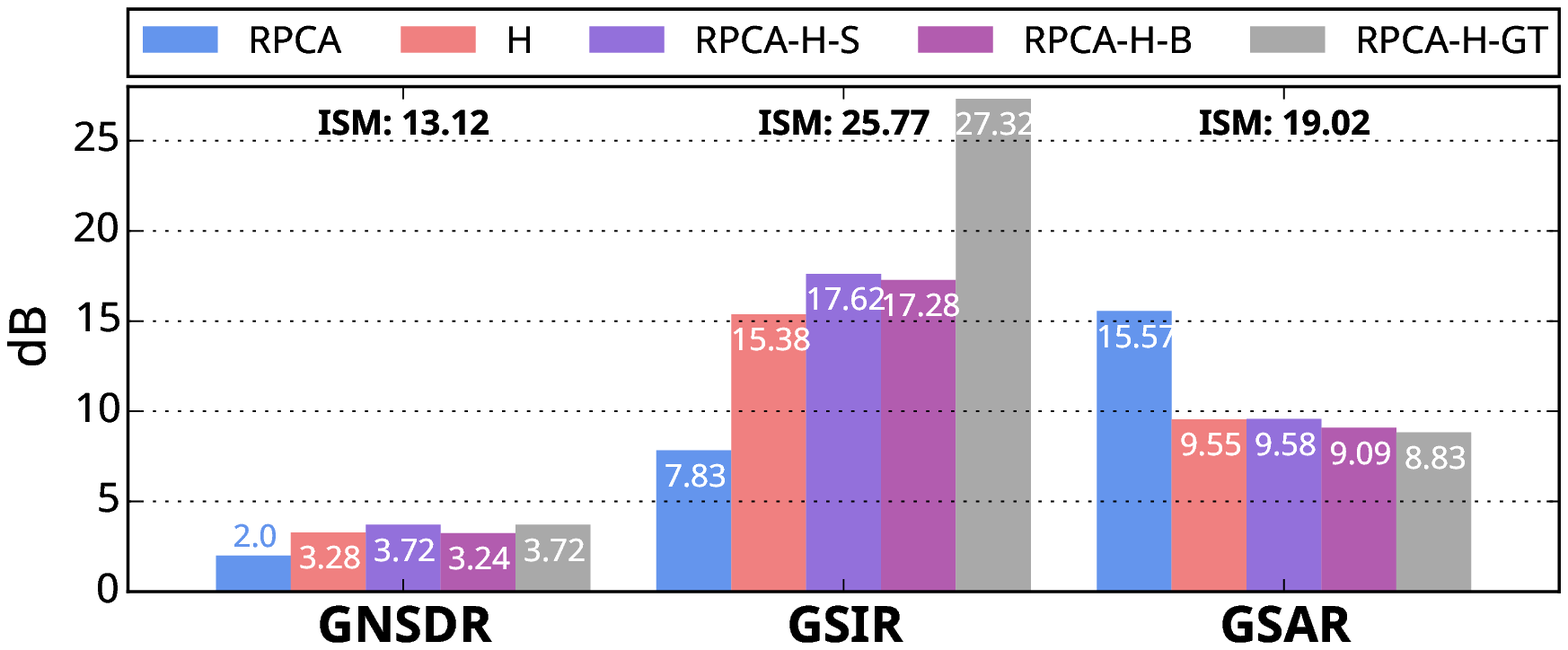}\\
      \vspace{-5pt}
      Accompaniment\\
      \vspace{-10pt}
      \includegraphics[width=0.99\columnwidth]{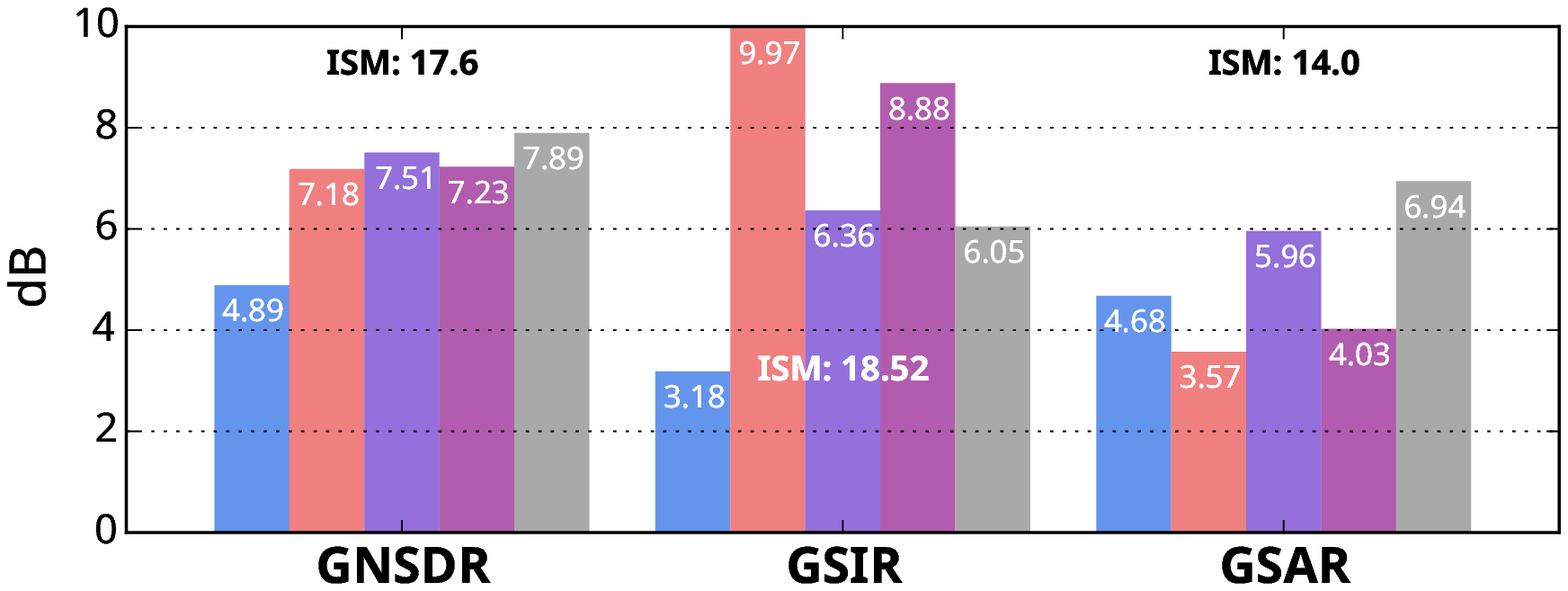}\\
      \text{(c) $5$ dB SNR}
    \end{minipage}
  \end{tabular}
  \caption{Comparative results of singing voice separation
    using different binary masks.
    The upper section shows the results for MIR-1K and the lower section for MedleyDB.
    From left to right, the results for mixing conditions at SNRs of $-$5, 0, and 5 dB are shown.
    \revise{The evaluation values of ``ISM'' are expressed with letters in order to make the graphs more readable.}
  \label{separation_result}}
\end{figure*}

%
% figure
%
\begin{figure*}[t]
  \centering
  \begin{minipage}{.19\textwidth}
    \centering
    \includegraphics[width=0.99\columnwidth]{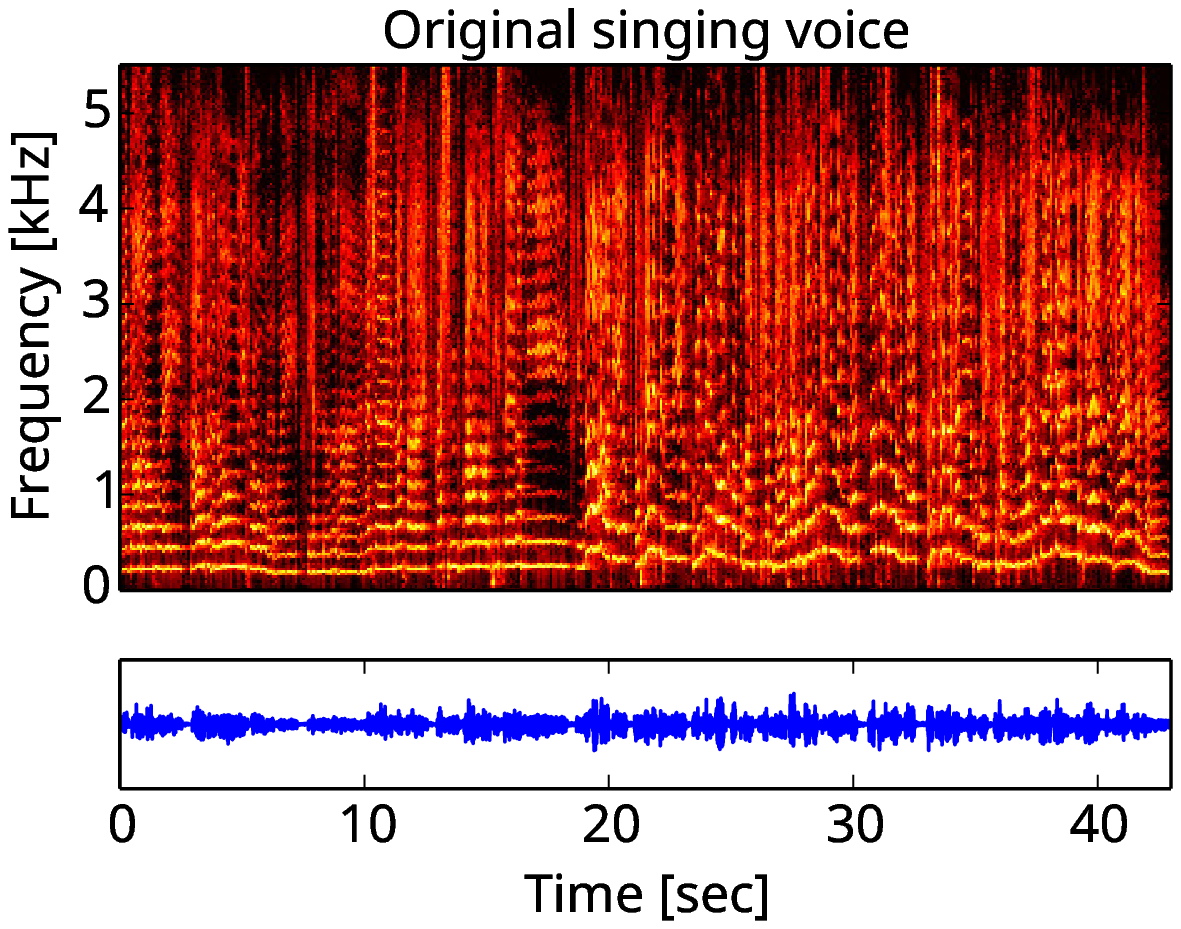}
  \end{minipage}
  \begin{minipage}{.19\textwidth}
    \centering
    \includegraphics[width=0.99\columnwidth]{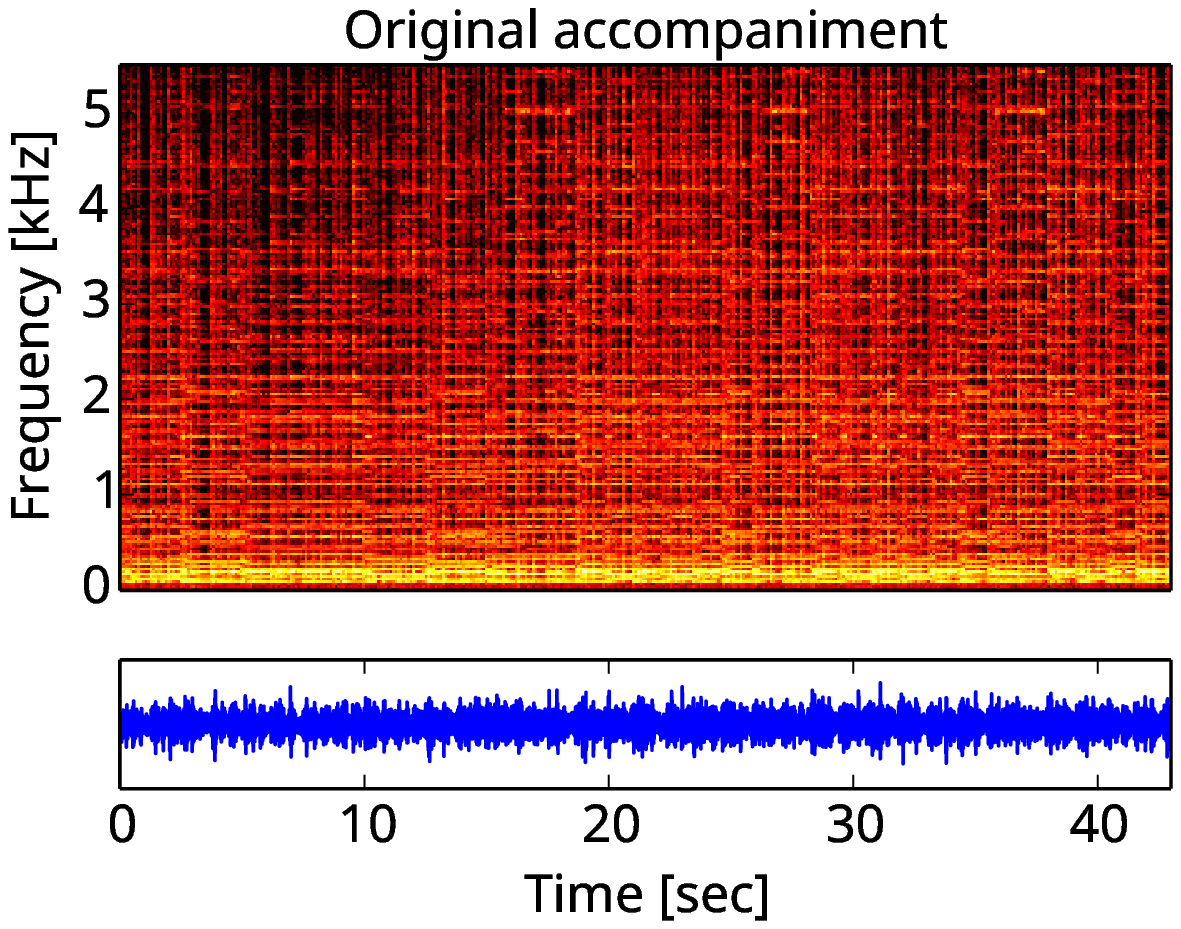}
  \end{minipage}
  \begin{minipage}{.19\textwidth}
    \centering
    \includegraphics[width=0.99\columnwidth]{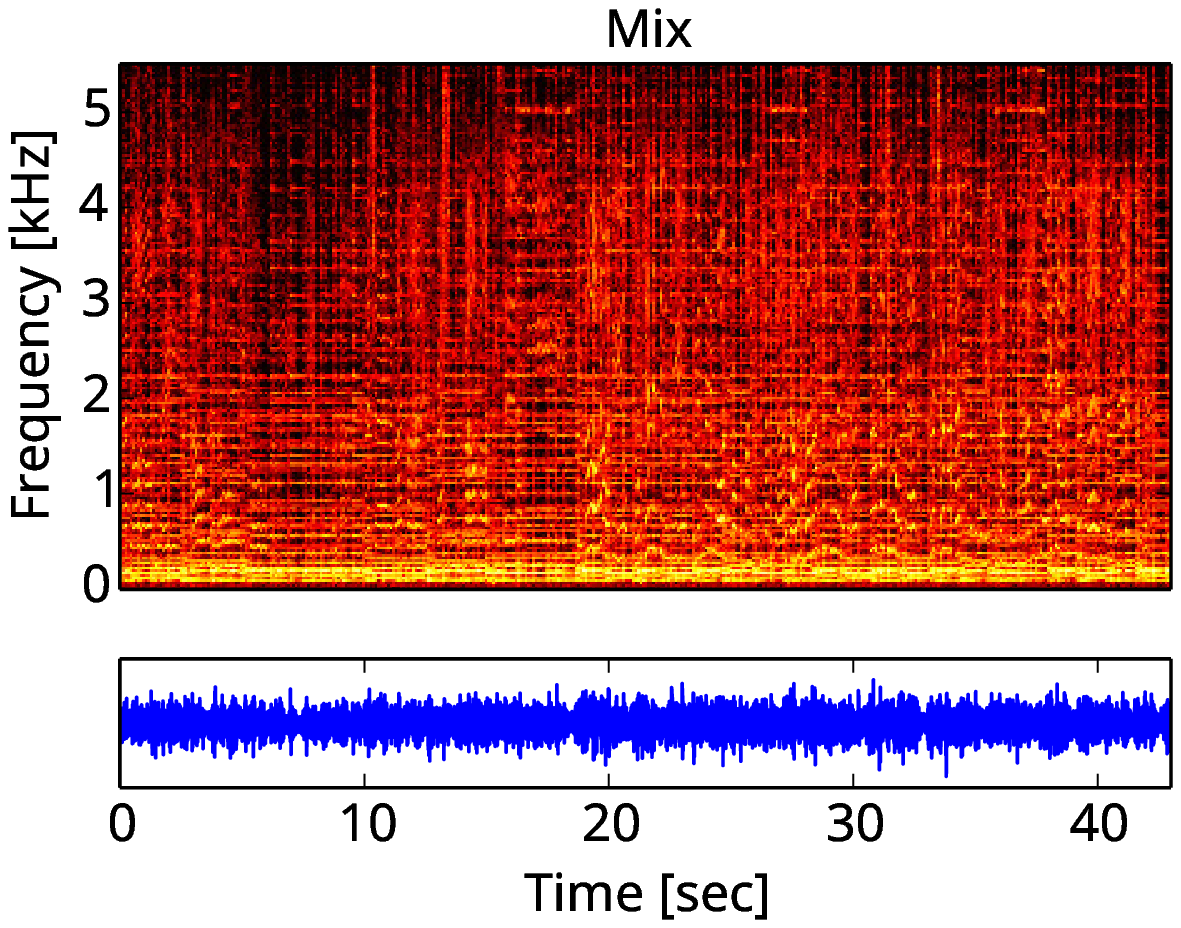}
  \end{minipage}
  \begin{minipage}{.19\textwidth}
    \centering
    \includegraphics[width=0.99\columnwidth]{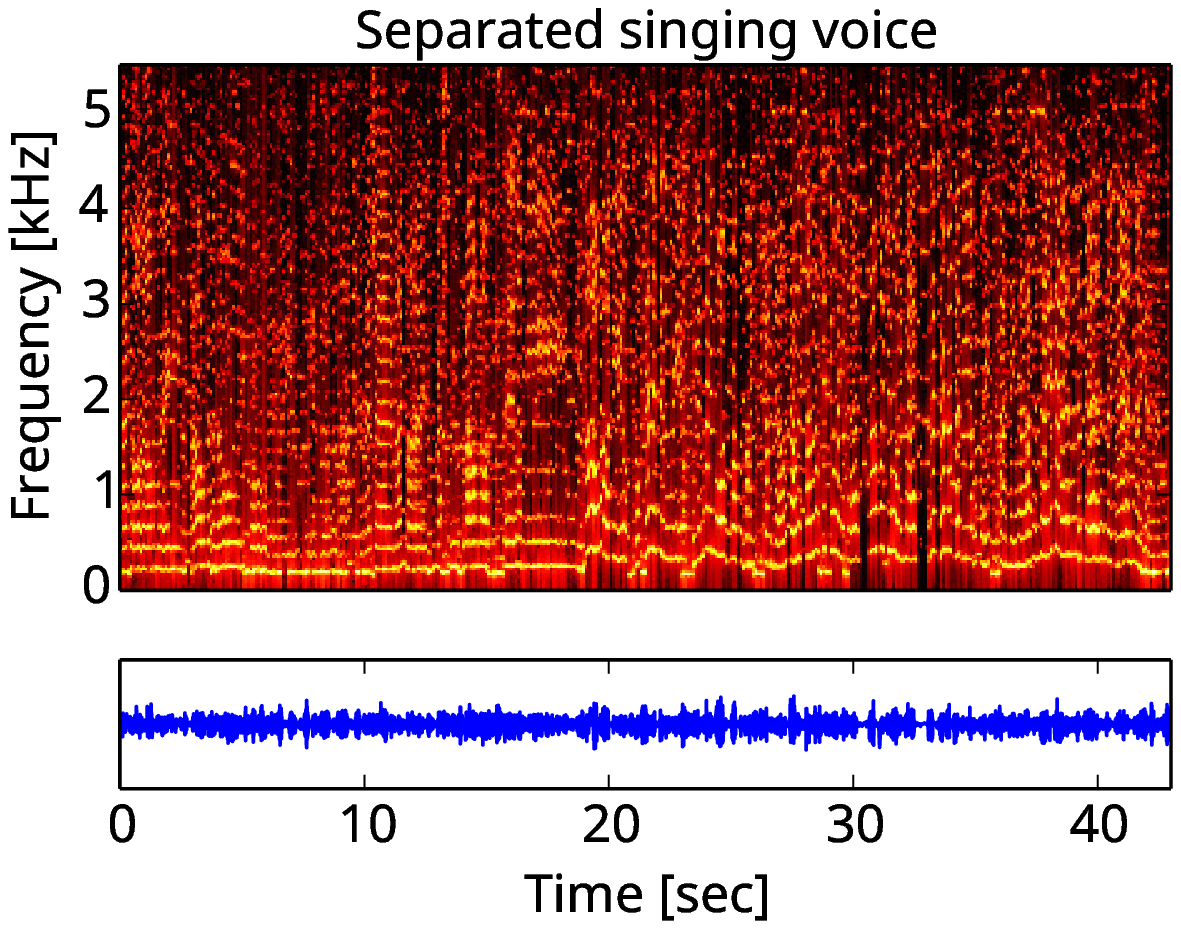}
  \end{minipage}
  \begin{minipage}{.19\textwidth}
    \centering
    \includegraphics[width=0.99\columnwidth]{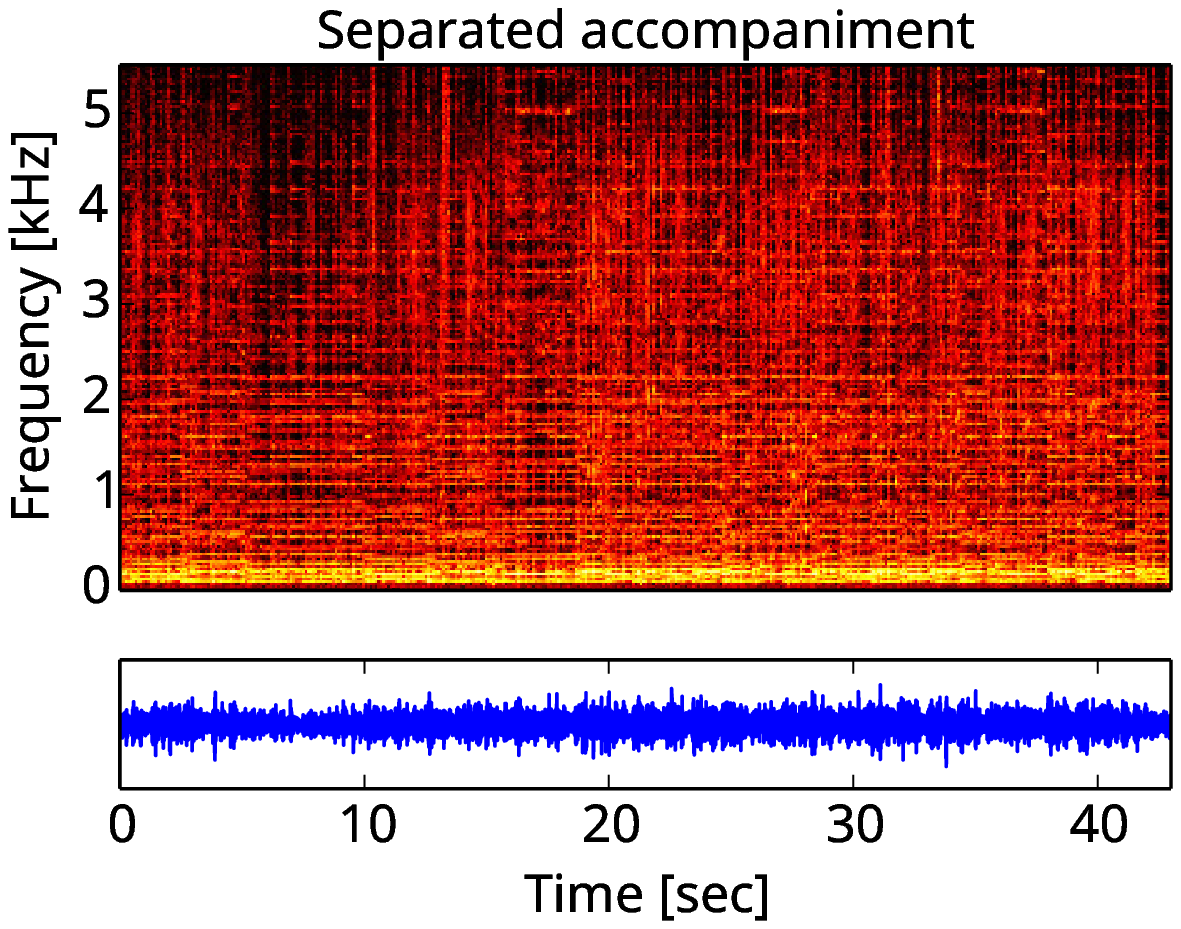}
  \end{minipage}
  \caption{An example of singing voice separation by the proposed method.
    The results of ``Coldwar / LizNelson'' in MedleyDB mixed at a $-$5 dB SNR are shown.
    From left to right, an original singing voice, an original accompaniment sound,
    a mixed sound, a separated singing voice, and a separated accompaniment sound are shown.
    The upper figures are spectrograms obtained by taking the STFT and the lower figures are resynthesized time signals.
    \label{sep_example}}
\end{figure*}

% \subsubsection{Experimental Conditions}

\subsubsection{Datasets and Parameters}

The MIR-1K dataset\footnote{https://sites.google.com/site/unvoicedsoundseparation/mir-1k} ({\it MIR-1K})
and the MedleyDB dataset ({\it MedleyDB}) \cite{medleydb} were used for evaluating singing voice separation.
Note that we used the 110 ``Undivided'' song clips of MIR-1K
and the 45 clips of MedleyDB listed in Table \ref{medleydb_used_clips}.
The clips in MIR-1K were recorded at a 16 kHz sampling rate with 16 bit resolution
and the clips in MedleyDB were recorded at a 44.1 kHz sampling rate with 16 bit resolution.
For each clip in both datasets,
singing voices and accompaniment sounds were mixed at three signal-to-noise ratios (SNR) conditions:
$-$5, 0, and 5 dB.

The datasets and the parameters used for evaluation are summarized
in Table \ref{tab:setting}, where
the parameters for computing 
the STFT (window size and hopsize), 
SHS (the number $N$ of harmonic partials), RPCA (a sparsity factor $\lambda$),
a harmonic mask (frequency width $w$),
and a saliency spectrogram (a weighting factor $\alpha$) are listed.
We empirically determined the parameters $w$ and $\lambda$ according to
the results of grid search (see details in Section \ref{system_parameters}).
The same value of $\lambda$ (0.8) was used for both RPCA computations in Fig.\ref{overview}.
The frequency range for the vocal F0 search was restricted to 80--720 Hz.

\subsubsection{Compared Methods}

The following binary masks were compared.
%% \begin{itemize}
%%   \setlength{\leftskip}{1.7cm}
%%   \item[{\bf RPCA}:] Using only an RPCA mask
%%   \item[{\bf H}:] Using only a harmonic mask
%%   \item[{\bf RPCA-H}:] Using an RPCA mask + a harmonic mask (\textbf{proposed method})
%%   \item[{\bf RPCA-H-GT}:] Using an RPCA mask + a harmonic mask made by using a ground-truth F0 contour
%%   \item[{\bf Ideal}:] Using an ideal binary mask
%% \end{itemize}
\revise{\begin{itemize}
  \setlength{\leftskip}{1.7cm}
  \item[{\bf RPCA}:] Using only an RPCA soft mask $\bm{M}_{\mbox{\tiny RPCA}}^{(\mathrm{s})}$
  \item[{\bf H}:] Using only a harmonic mask $\bm{M}_{\mbox{\tiny H}}$
  \item[{\bf RPCA-H-S}:] Using an integrated soft mask $\bm{M}_{\mbox{\tiny RPCA+H}}^{(\mathrm{s})}$ % (\textbf{proposed method})
  \item[{\bf RPCA-H-B}:] Using an integrated binary mask $\bm{M}_{\mbox{\tiny RPCA+H}}^{(\mathrm{b})}$ % (\textbf{proposed method})
  \item[{\bf RPCA-H-GT}:] Using an integrated soft mask made by using a ground-truth F0 contour
  \item[{\bf ISM}:] Using an ideal soft mask
\end{itemize}}%
``RPCA'' is a conventional RPCA-based method \cite{huang:2012}.
``H'' used only a harmonic mask created from an estimated F0 contour.
%% ``RPCA-H'' represents the proposed method, and
%% ``RPCA-H-GT'' represents the situation that vocal F0 estimation  performs perfectly
%% (the upper bound of separation quality for the proposed framework).
%% ``Ideal'' represents the situation that uses an ideal binary mask
%% (the upper bound of separation quality 
%% when using binary masking in the TF domain).
\revise{
``RPCA-H-S'' and ``RPCA-H-B'' represent
 the proposed methods using soft masks and binary masks, respectively, 
 and ``RPCA-H-GT'' means a condition that the ground-truth vocal F0s were given 
 (the upper bound of separation quality for the proposed framework).
``ISM'' represents a condition that oracle TF masks were estimated
 such that the ground-truth vocal and accompaniment spectrograms were obtained
 (the upper bound of separation quality of TF masking methods).
}%
For H, RPCA-H-S and RPCA-H-B,
the accuracies of vocal F0 estimation
are described in Section \ref{experiment_f0}.

\subsubsection{Evaluation Measures}

The {\it BSS\_EVAL} toolbox\footnote{http://bass-db.gforge.inria.fr/bss\_eval/} \cite{bss_eval}
was used for measuring the separation performance.
The principle of {\it BSS\_EVAL} is to decompose an estimate $\hat{s}$ of a true source signal $s$ as follows:
\begin{align}
  \hat{s}(t) = s_{\mathrm{target}}(t) + e_{\mathrm{interf}}(t) + e_{\mathrm{noise}}(t) + e_{\mathrm{artif}}(t),
\end{align}
where $s_{\mathrm{target}}$ is an allowed distortion of the target source $s$
and $e_{\mathrm{interf}}$, $e_{\mathrm{noise}}$ and $e_{\mathrm{artif}}$ are respectively the interference 
of the unwanted sources, perturbing noise, and artifacts in the separated signals
(such as musical noise).
Since we assume that an original signal consists of only vocal and accompaniment sounds,
the perturbing noise $e_{\mathrm{noise}}$ was ignored.
Given the decomposition, three performance measures are defined:
the Source-to-Distortion Ratio (SDR), the Source-to-Interference Ratio (SIR)
and the Source-to-Artifacts Ratio (SAR):
\begin{align}
  \mathrm{SDR}(\hat{s},s) &:= 10\log_{10}\left(\frac{\|s_{\mathrm{target}}\|^2}{\|e_{\mathrm{interf}} + e_{\mathrm{artif}}\|^2}\right), \\
  \mathrm{SIR}(\hat{s},s) &:= 10\log_{10}\left(\frac{\|s_{\mathrm{target}}\|^2}{\|e_{\mathrm{interf}}\|^2}\right), \\
  \mathrm{SAR}(\hat{s},s) &:= 10\log_{10}\left(\frac{\|s_{\mathrm{target}} + e_{\mathrm{interf}}\|^2}{\|e_{\mathrm{artif}}\|^2}\right),
\end{align}
where $\|\cdot\|$ denotes a Euclidean norm.
We then calculated the Normalized SDR (NSDR) that measures the improvement of the SDR between the estimate $\hat{s}$ 
of a target source signal $s$ and the original mixture $x$.
To measure the overall separation performance we calculated
the Global NSDR (GNSDR), 
which is a weighted mean of the NSDRs over all the mixtures $x_k$ (weighted by their length $l_k$):
\begin{align}
  \mathrm{NSDR}(\hat{s},s,x) &= \mathrm{SDR}(\hat{s},s) - \mathrm{SDR}(x,s), \\
  \mathrm{GNSDR} &= \frac{\sum_{k}l_k \mathrm{NSDR}(\hat{s}_k,s_k,x_k)}{\sum_{k}l_k}.
\end{align}
In the same way, the Global SIR (GSIR) and the Global SAR (GSAR) were calculated from the SIRs and the SARs.
For all these ratios, higher values represent better separation quality.

Since this paper does not deal with the VAD and we intended to examine the effect of the harmonic mask for vocal separation, 
we used only the voiced sections for evaluation; that is to say, the amplitude of the signals in unvoiced sections was set to 0 
when calculating the evaluation scores.

\subsubsection{Experimental Results}

Fig. \ref{separation_result} shows the evaluation results.
\revise{
In spite of F0 estimation errors,
 the proposed methods using soft masks (RPCA-H-S) and those using binary masks (RPCA-H-B) 
 outperformed both RPCA and H in GNSDR for all datasets. 
This indicates that
 combining an RPCA mask and a harmonic mask is effective 
 for improving the separation quality of singing voices and accompaniment sounds. 
The removal of the spectra of non-repeating instruments
 ({\it e.g.}, bass guitar) significantly improved the separation quality. 
RPCA-H-S outperformed RPCA-H-B in GNSDR, GSAR, and GSIR of the singing voice. 
On the other hand, RPCA-H-B outperformed RPCA-H-S in GSIR of the accompaniment
 and H outperformed both RPCA-H-B and RPCA-H-S. 
This indicates that a harmonic mask is useful for singing voice suppression.
}

Fig. \ref{sep_example} shows an example of an output 
of singing voice separation by the proposed method.
We can see that vocal and accompaniment sounds
were sufficiently separated from a mixed signal
even though the volume level of vocal sounds was lower than
that of accompaniment sounds.

%
% Table
%
\begin{table*}[t]
  \centering
  \caption{Experimental results for vocal F0 estimation
    (average accuracy [{\it \%}] over all clips in each dataset).
    \label{melody_result}}
  \begin{tabular}{c|c||ccccccc} \Hline
    \multicolumn{2}{c||}{} & \multicolumn{2}{c}{PreFEst-V} & \multicolumn{2}{c}{MELODIA-V} & \multicolumn{2}{c}{MELODIA} & {\bf Proposed} \\ \hline
    Database & SNR [dB] & w/o RPCA & w/ RPCA & w/o RPCA & w/ RPCA & w/o RPCA & w/ RPCA &  \\ \hline
    \multirow{3}{*}{MIR-1K} & $-$5 & 36.45 & 42.99 & 53.48 & {\bf 60.69} & 54.37 & 59.50 & 57.78  \\
     & 0 & 50.70 & 56.15 & 76.88 & {\bf 80.90} & 78.09 & 79.91 & 75.48  \\
    & 5 & 63.77 & 66.32 & 88.87 & {\bf 90.26} & 88.89 & 89.33 & 85.42  \\ \cline{1-2}
    \multirow{4}{*}{MedleyDB} & original mix & 70.83 & 72.25 & 70.69 & 74.93 & 71.24 & 73.40 & {\bf 81.90}  \\ 
    & $-$5 & 71.82 & 72.72 & 72.05 & 76.75 & 74.56 & 75.32 & {\bf 82.68}  \\ 
    & 0 & 80.91 & 81.02 & 86.59 & 89.20 & 87.34 & 87.54 & {\bf 90.31}  \\ 
    & 5 & 86.39 & 85.41 & 92.63 & {\bf 93.93} & 93.08 & 92.50 & 93.15 \\ \cline{1-2}
    RWC-MDB-P-2001 &  & 69.81 & 71.71 & 67.79 & 71.64 & 69.89 & 70.30 & {\bf 80.84}  \\ \Hline
    \multicolumn{2}{c||}{Average of all datasets} & 66.24 & 68.57 & 76.12 & 79.79 & 77.18 & 78.48 & {\bf 80.95}  \\ \Hline
  \end{tabular}
\end{table*}

\subsection{Vocal F0 Estimation}
\label{experiment_f0}

We compared the vocal F0 estimation of the proposed method with conventional methods.

\subsubsection{Datasets}

MIR-1K, MedleyDB, and the RWC Music Database ({\it RWC-MDB-P-2001}) \cite{rwc}
were used for evaluating vocal F0 estimation.
RWC-MDB-P-2001 contains 100 song clips of popular music
which were recorded at a 44.1 kHz sampling rate with 16 bit resolution.
The dataset contains 20 songs with English lyrics performed in the style of American popular music in the 1980s 
and 80 songs with Japanese lyrics performed in the style of Japanese popular music in the 1990s.

\subsubsection{Compared Methods}

The following four methods were compared.
\begin{itemize}
  \setlength{\leftskip}{1.8cm}
  \item[{\bf PreFEst-V}:] PreFEst (saliency spectrogram) + Viterbi search
  \item[{\bf MELODIA-V}:] MELODIA (saliency spectrogram) + Viterbi search
  \item[{\bf MELODIA}:] The original MELODIA algorithm
  \item[{\bf Proposed}:] F0-saliency spectrogram + Viterbi (\textbf{proposed method})
\end{itemize}
{\it PreFEst} \cite{goto:2004} is a statistical multi-F0 analyzer
that is still considered
to be competitive for vocal F0 estimation.
 Although PreFEst contains three processes ---the {\it PreFEst-front-end} for frequency
analysis, the {\it PreFEst-core} computing a saliency spectrogram, 
and the {\it PreFEst-back-end} 
that tracks F0 contours using multiple agents
---we used only the {\it PreFEst-core} and estimated F0 contours
by using the Viterbi search described in Section \ref{viterbi_search} (``PreFEst-V''). 
{\it MELODIA} is a state-of-the-art algorithm for vocal F0 estimation that 
focuses on the characteristics of vocal F0 contours.
We applied the Viterbi search to a saliency spectrogram derived from MELODIA (``MELODIA-V'')
and also tested the original MELODIA algorithm (``MELODIA'').
In this experiment we used the MELODIA implementation provided as 
a vamp plug-in\footnote{http://mtg.upf.edu/technologies/melodia}.

Singing voice separation based on RPCA \cite{huang:2012} 
was applied before computing conventional methods as preprocessing
(``w/ RPCA'' in Table \ref{melody_result}).
We investigated the effectiveness of the proposed method
in conjunction with 
preprocessing of singing voice separation.

\subsubsection{Evaluation Measures}

We measured
the raw pitch accuracy (RPA) 
defined as the ratio of the number of frames
 in which correct vocal F0s were detected
 to the total number of voiced frames.
An estimated value was considered correct if
the difference between it and the ground-truth F0 
was 50 cents (half a semitone) or less.

\subsubsection{Experimental Results}

Table \ref{melody_result} shows the experimental results of vocal F0 estimation,
where each value is an average accuracy over all clips.
The results show that the proposed method achieved the best performance in terms of average accuracy.
With MedleyDB and RWC-MDB-P-2001
the proposed method significantly outperformed the other methods,
while the performance of
MELODIA-V and MELODIA were better than that of the proposed method
with MIR-1K.
% This might be due to 
% the tendency of instrument composition of each datasets.
\revise{
This might be due to the different instrumentation of songs included in each dataset. 
}%
Most clips in MedleyDB and RWC-MDB-P-2001
contain the sounds of many kinds of musical instruments, whereas 
most clips in MIR-1K
contain the sounds of only a small number of musical instruments.
%% It is difficult for the proposed method to distinguish
%% vocal sounds and melodic accompaniment sounds.

\revise{
These results are originated from the characteristics of the proposed method.
In vocal F0 estimation,
 the spectral periodicity of an RPCA binary mask is used to enhance vocal spectra.
The harmonic structures of singing voices appear clearly in the RPCA mask 
 when music audio signals contain various kinds of repetitive musical instrument sounds.
The proposed method therefore works well especially
for songs of particular genres such as {\it rock} and {\it pops}.
}\noindent

%
% Table
%
\begin{table}[t]
  \centering
  \caption{Parameter settings for MIREX2014.}
  \label{tab:setting:mirex}
  \begin{tabular}{c|ccccc}
  \Hline
     & Window size & Hopsize & $N$ & $\lambda$ & $w$ \\ \hline
    IIY1 & 4096 & 441 & 15 & 1.0 & 100 \\
    IIY2    & 4096 & 441 & 15 & 0.8 & 100 \\
  \Hline
  \end{tabular}
\end{table}

\subsection{MIREX2014}

We submitted our algorithm to the {\it Singing Voice Separation}
 task of the Music Information Retrieval Evaluation eXchange (MIREX) 2014,
 which is a community-based framework for the formal evaluation of analysis algorithms.
Since the datasets are not freely distributed to the participants, 
 MIREX provides meaningful and fair scientific evaluations.

There is some difference between our submission for MIREX and the algorithm described in this paper.
The major difference is that
% only an SHS spectrogram (excepting an F0 enhancement spectrogram in Section \ref{salience_f0est}) 
only an SHS spectrogram (\revise{with the exception of} an F0 enhancement spectrogram in Section \ref{salience_f0est}) 
was used as a saliency spectrogram in the submission.
Instead a simple vocal activity detection (VAD) method based on an energy threshold
was used after singing voice separation.

%
% figure
%
\begin{figure*}[t]
  \centering
  \begin{tabular}{c}
    \centering
    \begin{minipage}{.33\textwidth}
      \centering
      \includegraphics[width=0.99\columnwidth]{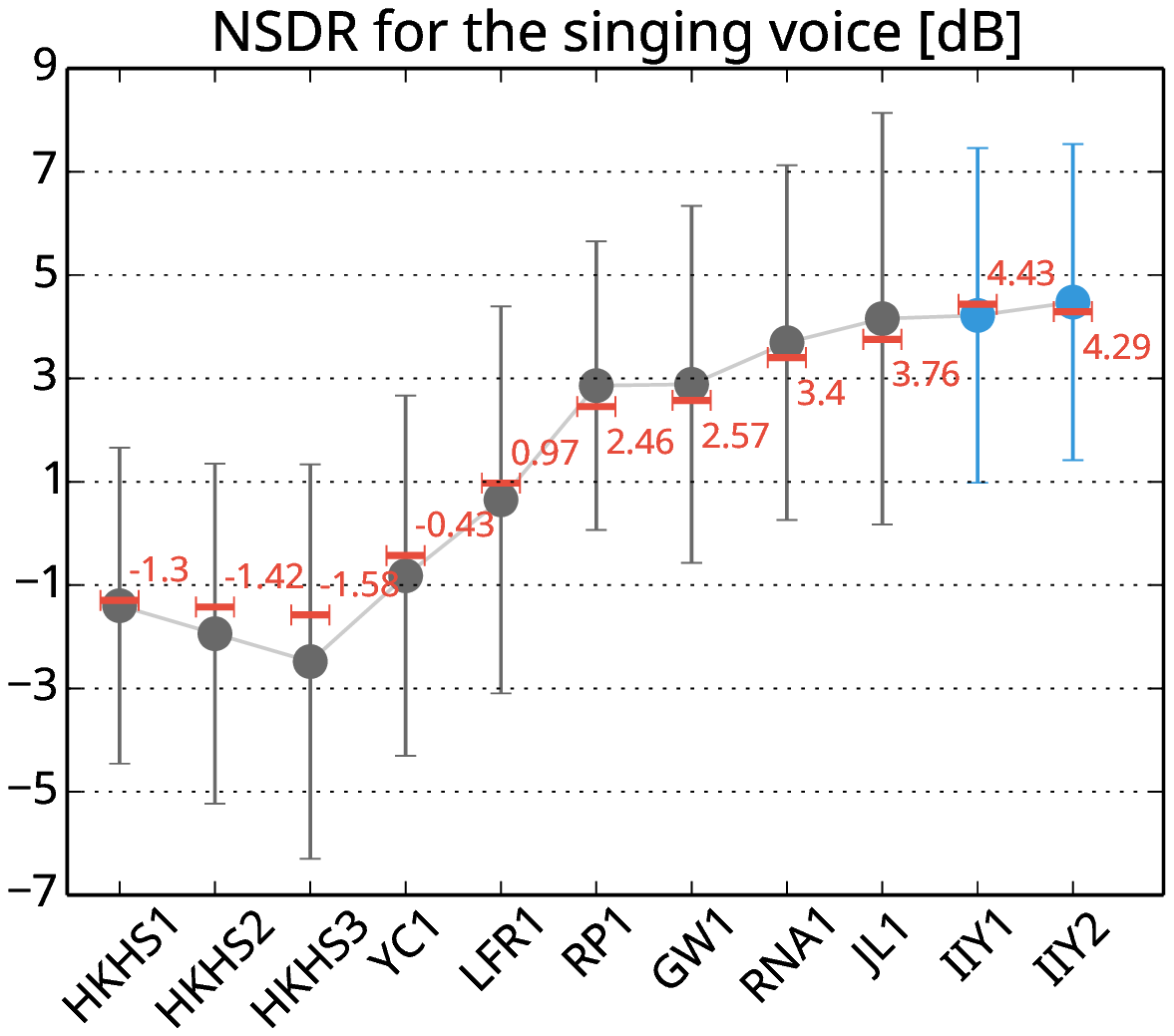}
    \end{minipage}
    \begin{minipage}{.33\textwidth}
      \centering
      \includegraphics[width=0.99\columnwidth]{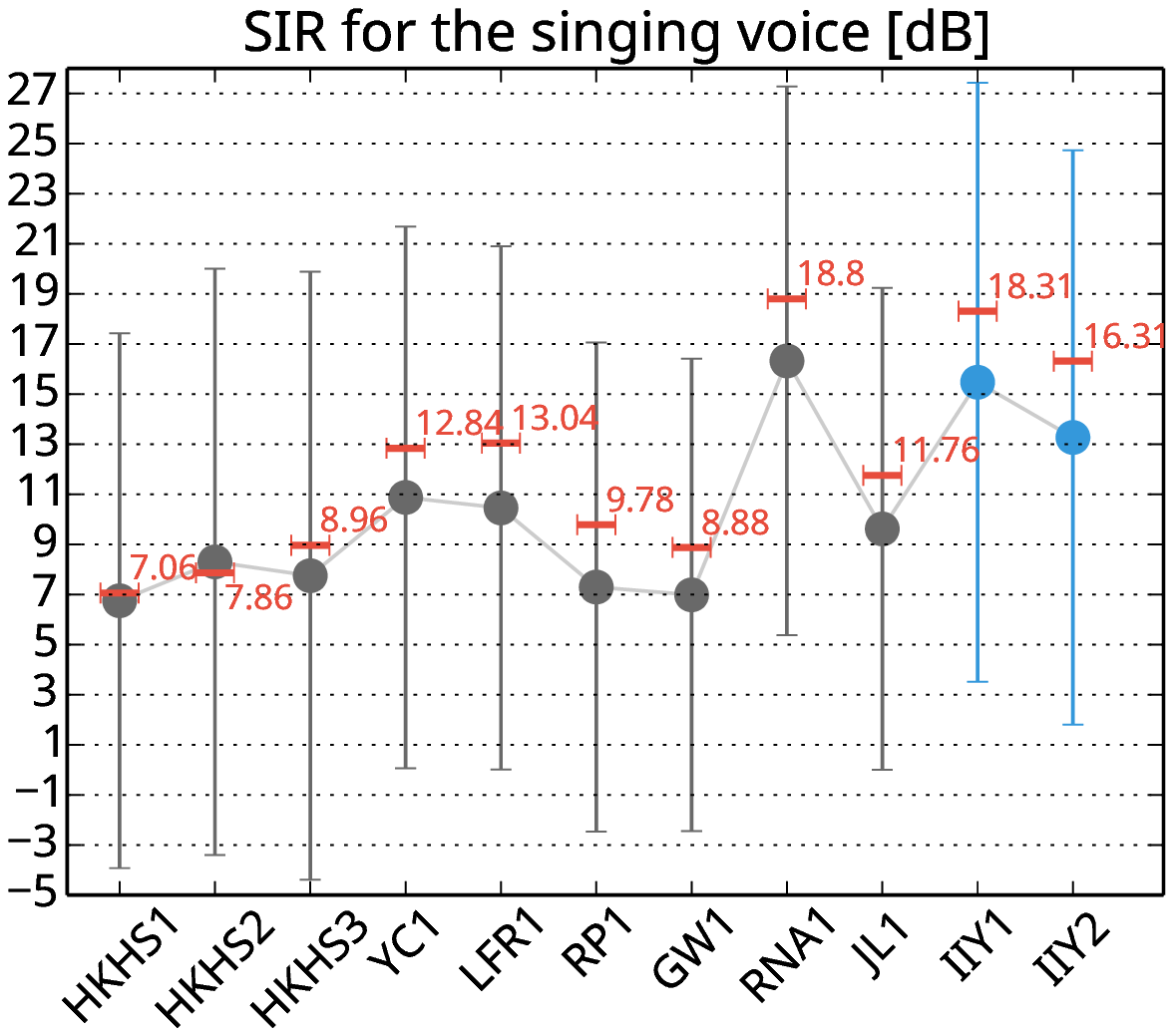}
    \end{minipage}
    \begin{minipage}{.33\textwidth}
      \centering
      \includegraphics[width=0.99\columnwidth]{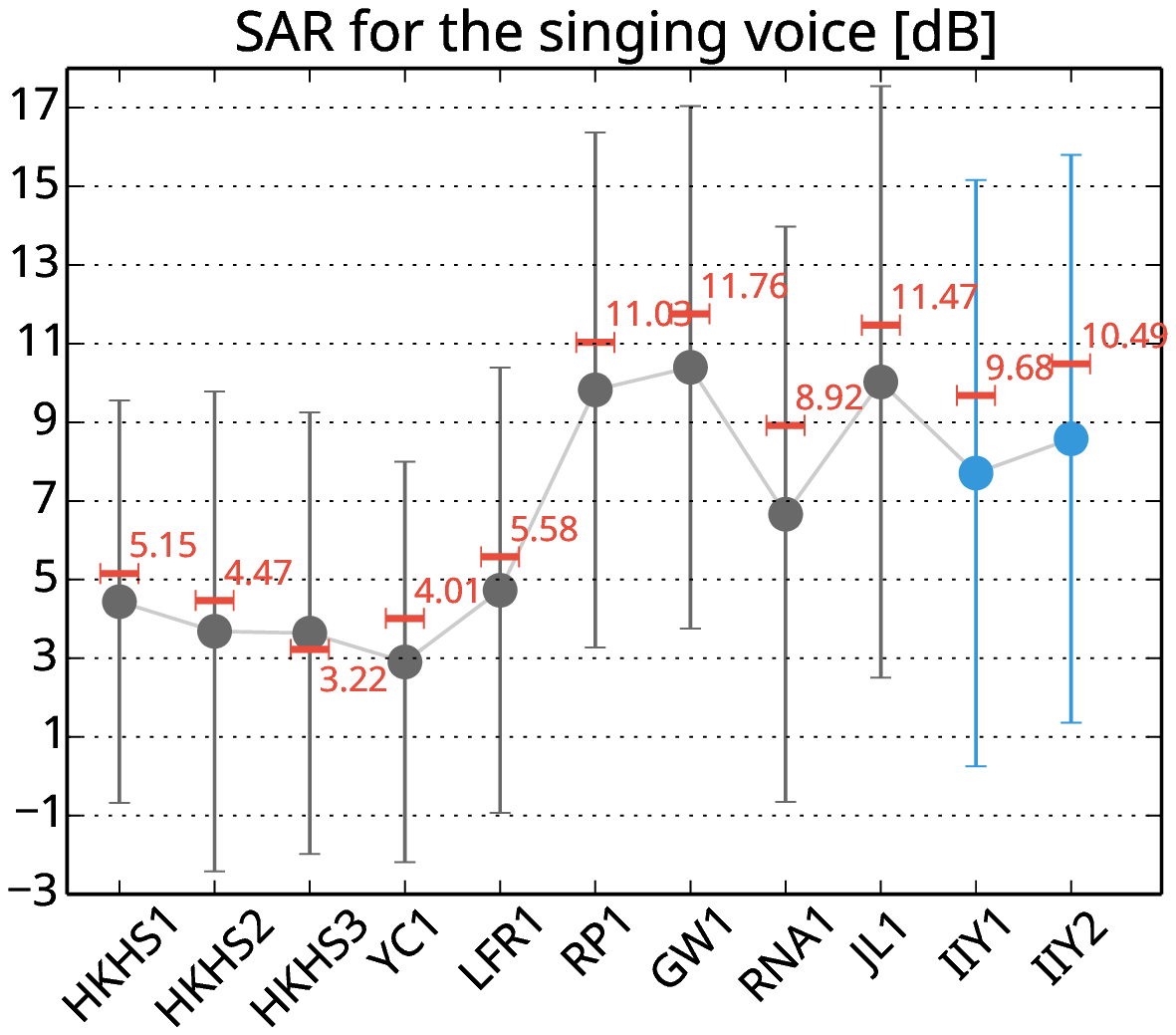}
    \end{minipage}
  \end{tabular}
  \begin{tabular}{c}
    \begin{minipage}{.33\textwidth}
      \centering
      \includegraphics[width=0.99\columnwidth]{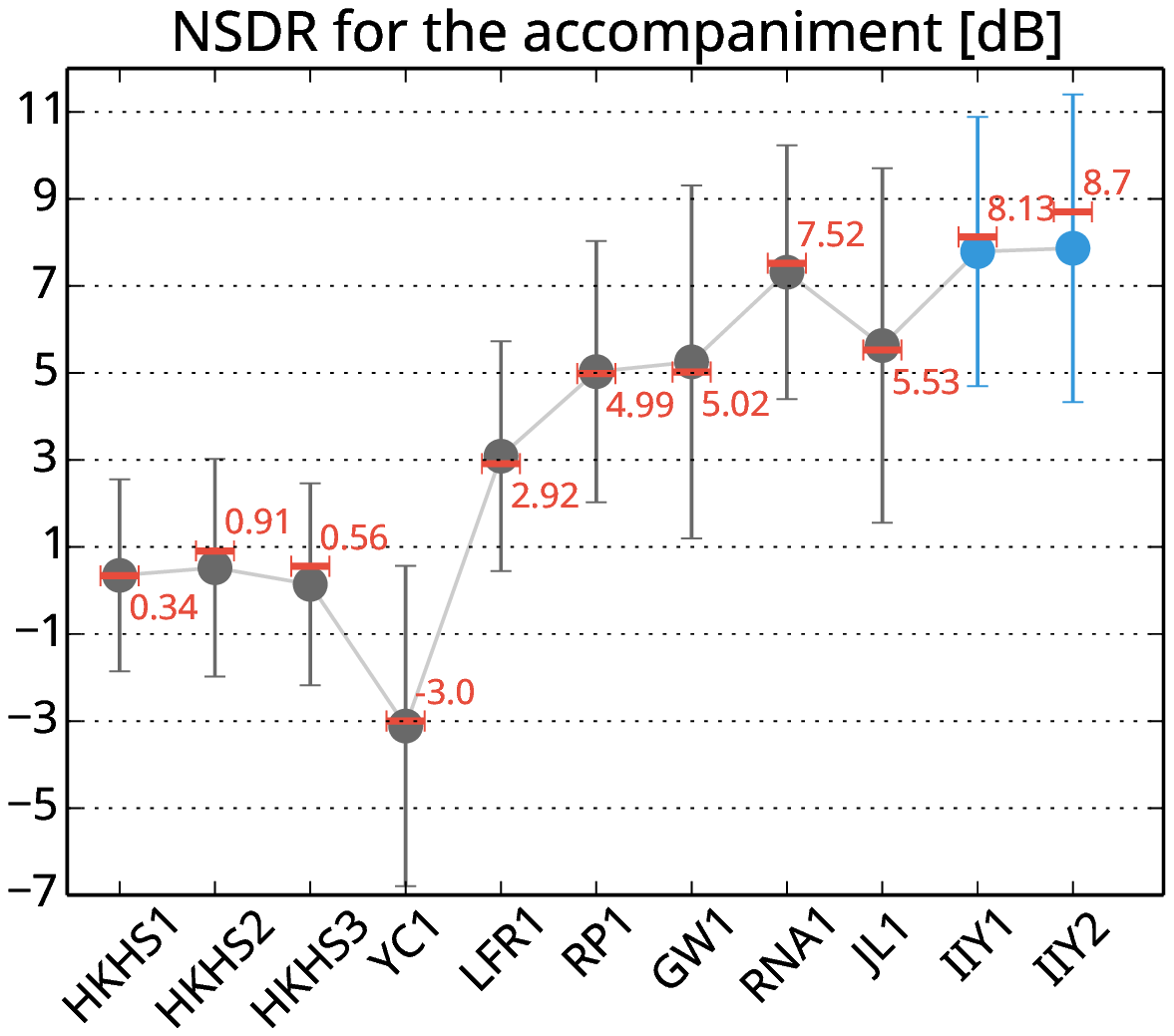}
    \end{minipage}
    \begin{minipage}{.33\textwidth}
      \centering
      \includegraphics[width=0.99\columnwidth]{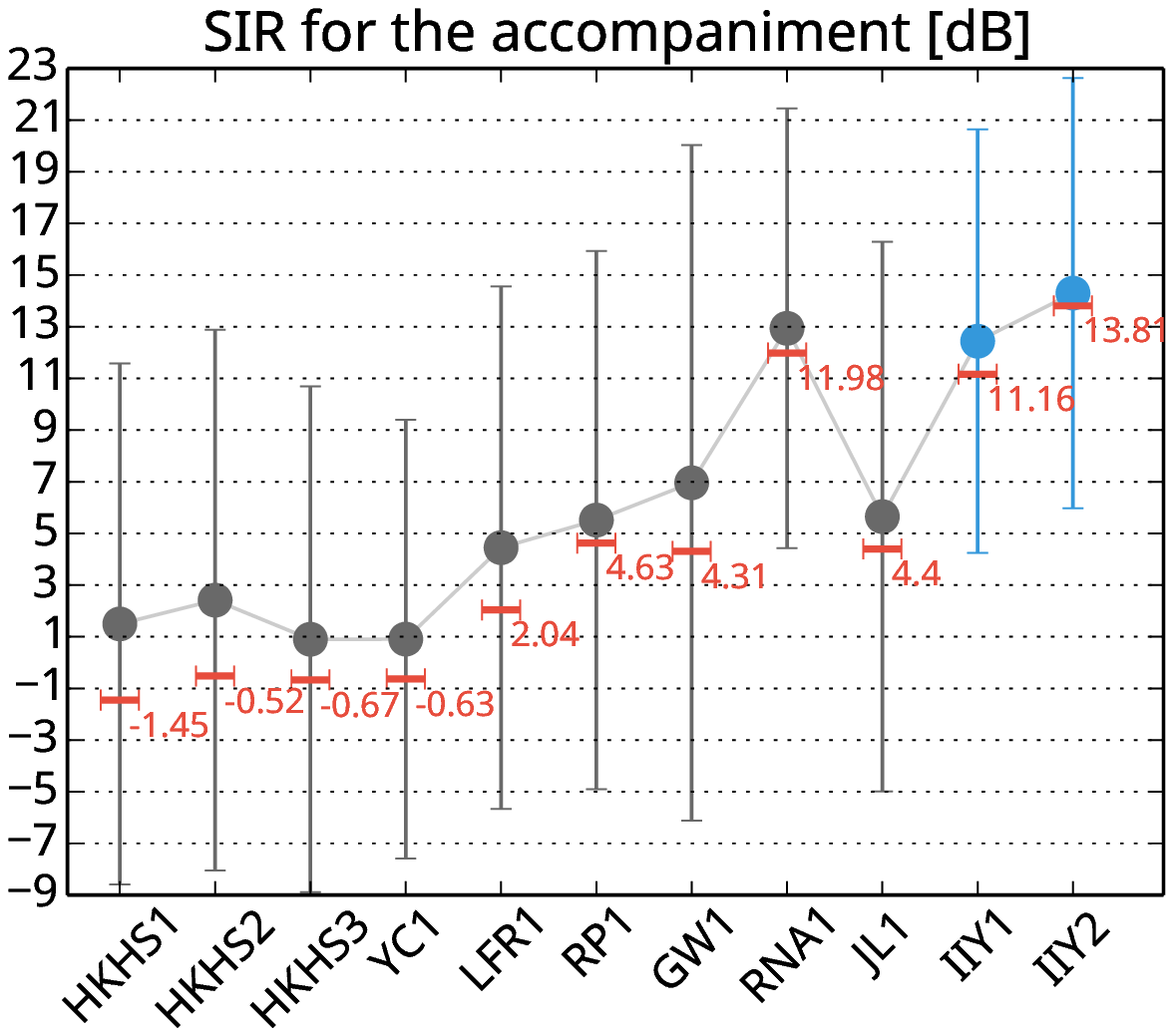}
    \end{minipage}
    \begin{minipage}{.33\textwidth}
      \centering
      \includegraphics[width=0.99\columnwidth]{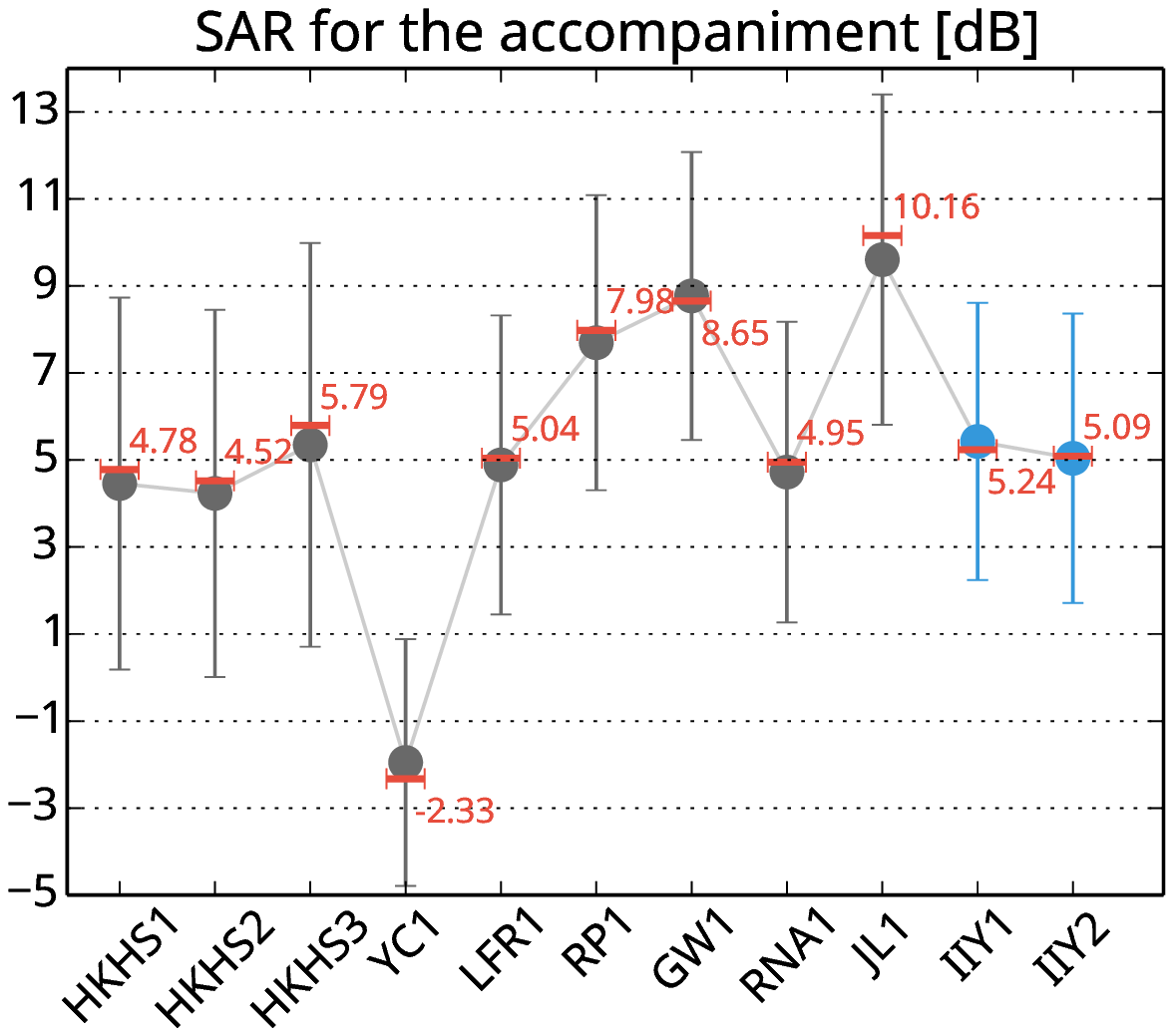}
    \end{minipage}
  \end{tabular}
  \caption{Results of the {\it Singing Voice Separation} task in MIREX2014.
  The circles, error bars, and red values represent means, standard deviations, 
  and medians for all song clips, respectively.
  \label{mirex_result}}
  \vspace{-10pt}
\end{figure*}

\subsubsection{Dataset}
100 monaural clips of pop music recorded at 44.1-kHz sampling rate with 16-bit resolution
were used for evaluation.
The duration of each clip was 30 seconds.

\subsubsection{Compared Methods}
11 submissions participated 
in the task\footnote{www.music-ir.org/mirex/wiki/2014:Singing\_Voice\_Separation\_Results}.
The submissions {\bf HKHS1}, {\bf HKHS2} and {\bf HKHS3} are algorithms
using deep recurrent neural networks~\cite{huang:2014}.
{\bf YC1} separates singing voices by clustering modulation features~\cite{yen:2014}.
{\bf RP1} is the REPET-SIM algorithm that identifies repetitive structures in polyphonic music 
by using a similarity matrix~\cite{repet_sim}.
{\bf GW1} uses Bayesian NMF to model a polyphonic spectrogram,
and clusters the learned bases based on acoustic features~\cite{yang:2014}.
{\bf JL1} uses the temporal and spectral discontinuity of singing voices~\cite{jeong:2014},
and {\bf LFR1} uses light kernel additive modeling based on the algorithm in~\cite{liutkus:2014}.
{\bf RNA1} first estimates predominant F0s
and then reconstructs an isolated vocal signal 
based on harmonic sinusoidal modeling using estimated F0s.
{\bf IIY1} and {\bf IIY2} are our submissions.
The only difference between IIY1 and IIY2 is their parameters.
The parameters for both submissions are listed in Table \ref{tab:setting:mirex}.

\subsubsection{Evaluation Results}
Fig. \ref{mirex_result} shows the evaluation results for all submissions.
Our submissions (IIY1 and IIY2) provided the best mean NSDR for both vocal and accompaniment sounds.
Even though the submissions using the proposed method outperformed the state-of-the-art methods in MIREX 2014,
there is still room for improving their performances.
As described in Section \ref{gridsearch_sep}, 
the robust range for the parameter $w$ is from 40 to 60.
We set the parameter to 100 in the submissions, however, and that
must have considerably reduced the sound quality of both separated vocal and accompaniment sounds.

\section{Parameter Tuning}
\label{system_parameters}

In this section
we discuss the effects of
parameters that determine the performances of singing voice separation and vocal F0 estimation.

%
% figure
%
\begin{figure}[t]
  \centering
  \begin{tabular}{c}
    $-$5 dB SNR
    \vspace{3pt}\\
    \begin{minipage}{0.48\columnwidth}
      \centering
      \footnotesize{GNSDRs for the singing voice}
      \includegraphics[width=\columnwidth]{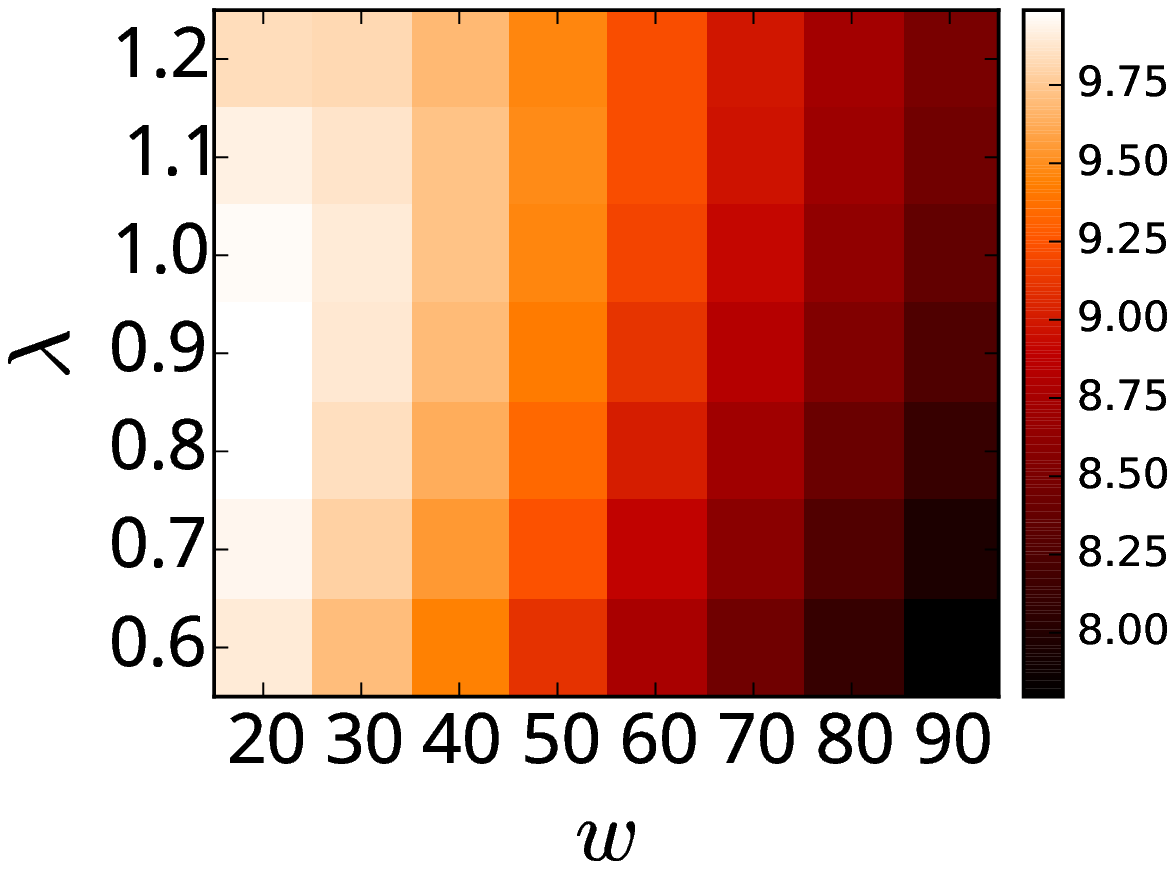}
    \end{minipage}
    \begin{minipage}{0.48\columnwidth}
      \hspace{-5pt}
      \centering
      \footnotesize{GNSDRs for the accompaniment}
      \includegraphics[width=\columnwidth]{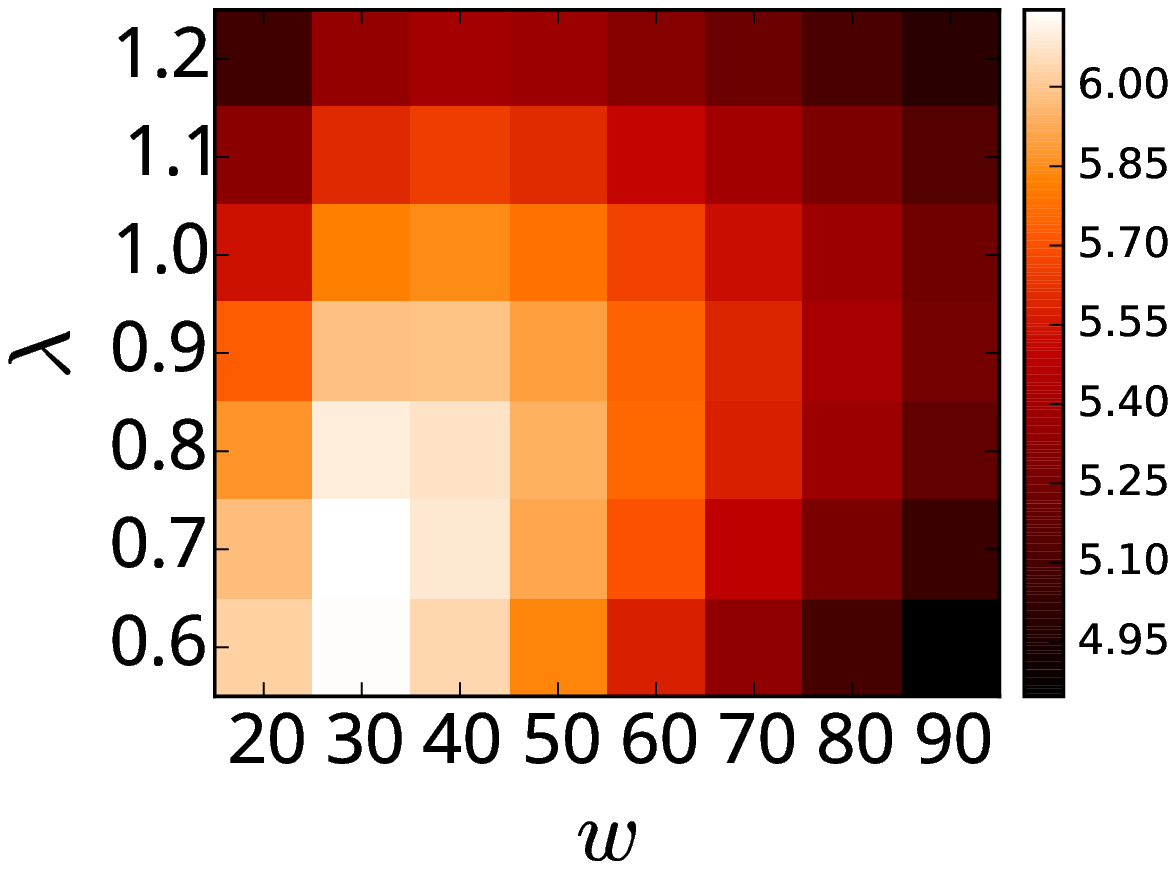}
    \end{minipage}
  \end{tabular}
  \begin{tabular}{c}
    0 dB SNR 
    \vspace{3pt}\\
    \begin{minipage}{0.48\columnwidth}
      \centering
      \footnotesize{GNSDRs for the singing voice}
      \includegraphics[width=\columnwidth]{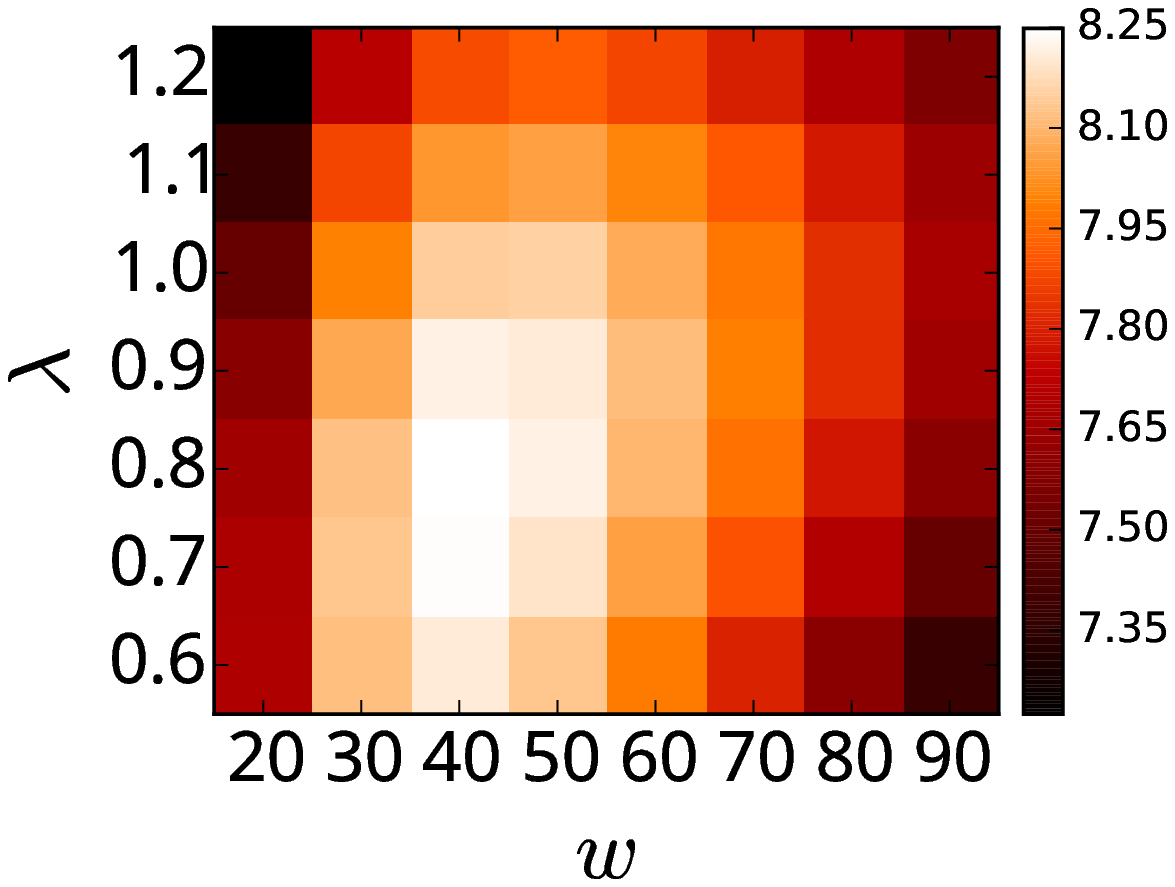}
    \end{minipage}
    \begin{minipage}{0.48\columnwidth}
      \hspace{-5pt}
      \centering
      \footnotesize{GNSDRs for the accompaniment}
      \includegraphics[width=\columnwidth]{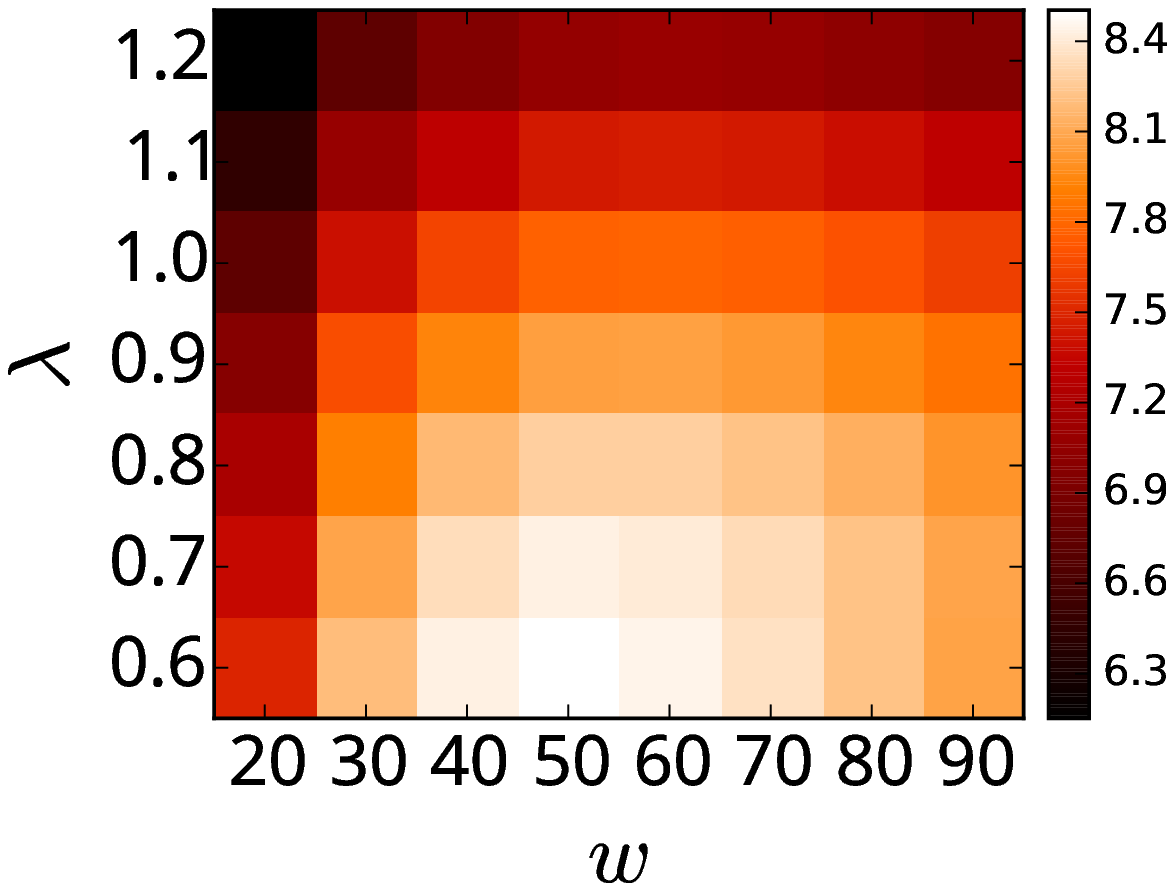}
    \end{minipage}
  \end{tabular}
  \begin{tabular}{c}
    5 dB SNR
    \vspace{3pt}\\
    \begin{minipage}{0.48\columnwidth}
      \centering
      \footnotesize{GNSDRs for the singing voice}
      \includegraphics[width=\columnwidth]{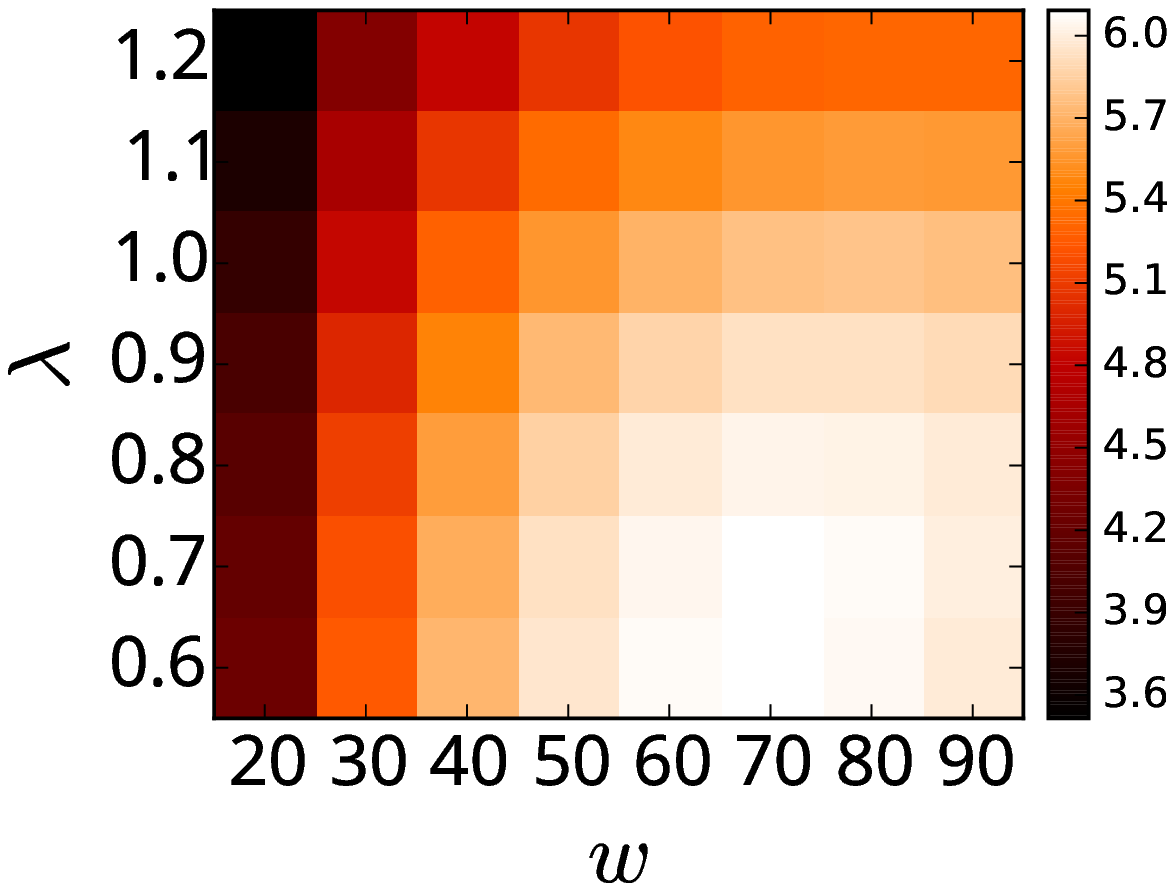}
    \end{minipage}
    \begin{minipage}{0.48\columnwidth}
      \hspace{-5pt}
      \centering
      \footnotesize{GNSDRs for the accompaniment}
      \includegraphics[width=\columnwidth]{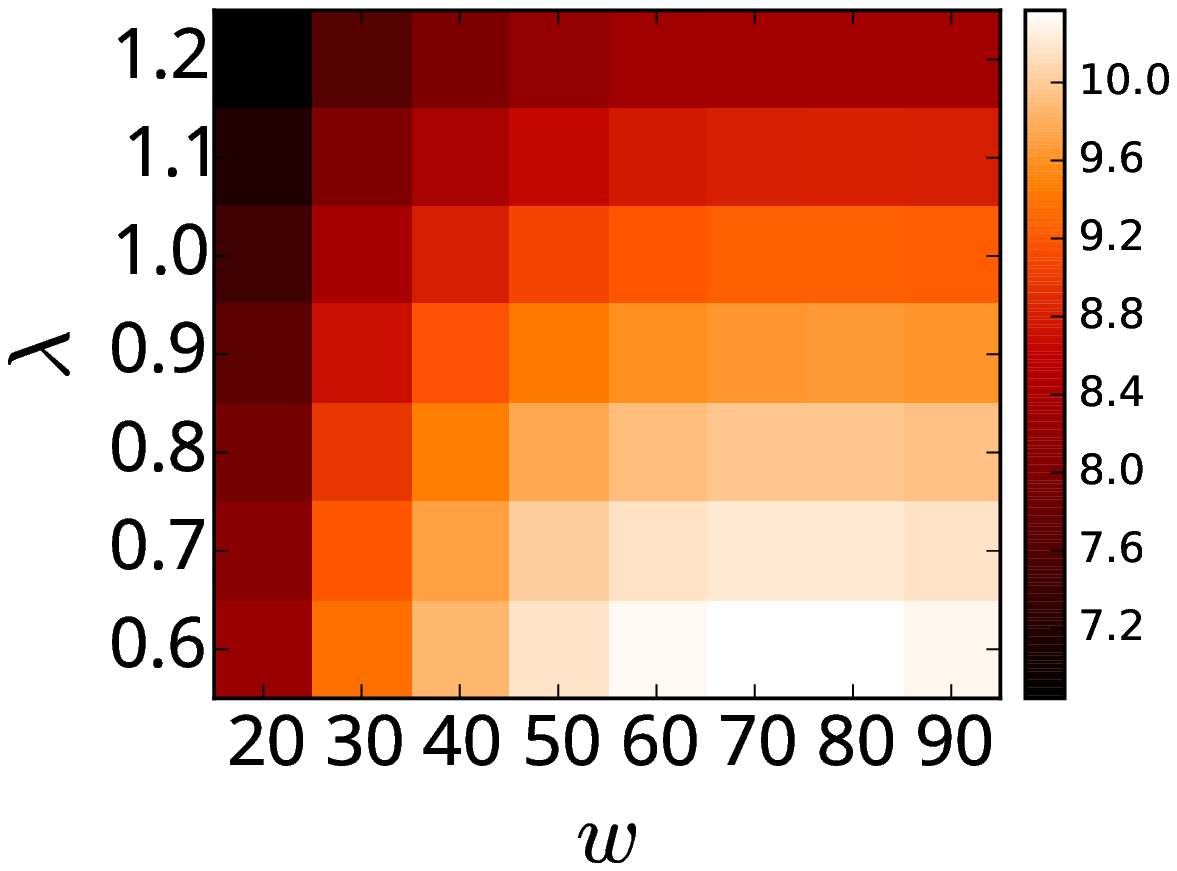}
    \end{minipage}
  \end{tabular}
  \caption{Experimental results of grid search for singing voice separation.
    GNSDR for MIR-1K is shown in each unit.
    From top to bottom, the results of $-$5, 0, and 5 dB SNR conditions are shown.
    The left figures show results for the singing voice and the right figures for the music accompaniment.
    In all parts of this figure, lighter values represent better results.
  \label{sep_para_grid_result}}
  \vspace{-10pt}
\end{figure}

\subsection{Singing Voice Separation}
\label{gridsearch_sep}

The parameters $\lambda$ and $w$ affect the quality of singing voice separation.
$\lambda$ is the sparsity factor of RPCA described in Section \ref{rpca-based method}
and $w$ is the frequency width of the harmonic mask described in Section \ref{harmonic_mask}.
The parameter $\lambda$ can be used to
trade off the rank of a low-rank matrix with the sparsity of a sparse matrix.
The sparse matrix is sparser
when $\lambda$ is larger and is less sparse when $\lambda$ is smaller.
When $w$ is smaller, 
fewer spectral bins around an F0 and its harmonic partials 
are assigned as singing voices.
This is the recall-precision trade-off of singing voice separation.
To examine the relationship between $\lambda$ and $w$,
we evaluated the performance of singing voice separation
for combinations of $\lambda$ from 0.6 to 1.2 in steps of 0.1
and $w$ from 20 to 90 in steps of 10.

\subsubsection{Experimental Conditions}

MIR-1K was used for evaluation
at three mixing conditions 
with SNRs of $-$5, 0, and 5 dB.
In this experiment, a harmonic mask was created using a ground-truth F0 contour
to examine only the effects of $\lambda$ and $w$.
GNSDRs were calculated for each parameter combination.

\subsubsection{Experimental Results}

Fig. \ref{sep_para_grid_result} shows
the overall performance for all parameter combinations.
Each unit on a grid represents the GNSDR value.
It was shown that $\lambda$ from 0.6 to 1.0 and $w$ from 40 to 60 provided robust performance 
in all mixing conditions.
In the $-$5 dB mixing condition,
an integrated mask performed better 
for both of the singing voice and the accompaniment
when $w$ was smaller.
This was because
most singing voice spectra were covered by accompaniment spectra
and only few singing voice spectra were dominant
around an F0 and harmonic partials
in the condition.

%\revise{Fig. \ref{sep_grid_1}} shows the experimental results of all performance measures
%for the $-$5 dB mixing condition when $w$ was 30.
%The SIR values for singing voices and those for accompaniment sounds had a trade-off relation
%in the proposed method.
%This was because $P_{\mathrm{amp}}$ increased and $R_{\mathrm{amp}}$ decreased
%when an RPCA mask and a harmonic mask were combined.

\subsection{Vocal F0 Estimation}

The parameters $\lambda$ and $\alpha$ affect the accuracy of vocal F0 estimation.
$\lambda$ is the sparsity factor of RPCA
and $\alpha$ is the weight parameter for computing the F0-saliency spectrogram described in Section \ref{salience_f0est}.
$\alpha$ determines the balance between an SHS spectrogram and an F0 enhancement spectrogram in a F0-saliency spectrogram,
and there must be range of its value
that provides robust performance.
We evaluated the accuracy of singing voice separation 
for combinations of $\lambda$ from 0.6 to 1.1 in steps of 0.1
and $\alpha$ from 0 to 2.0 in steps of 0.2.
RWC-MDB-P-2001 was used for evaluation,
and RPA was measured for each parameter combination.

\begin{figure}[t]
  \centering
  \includegraphics[width=0.9\columnwidth]{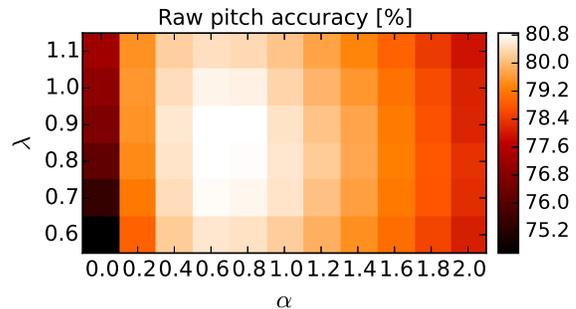}
  \caption{Experimental results of grid search for vocal F0 estimation.
  The mean raw pitch accuracy for RWC-MDB-P-2001 is shown in each unit.
  Lighter values represent better accuracy.
  \label{f0_para_grid_result}}
\end{figure}

Fig. \ref{f0_para_grid_result} shows
the overall performance for all parameter combinations of grid search.
Each unit on a grid represents RPA
for each parameter combination.
It was shown that $\lambda$ from 0.7 to 0.9 and $\alpha$ from 0.6 to 0.8 provided comparatively better performance 
than any other parameter combinations.
RPCA with $\lambda$ within the range separates vocal sounds to a moderate degree
for vocal F0 estimation.
The value of $\alpha$ was also crucial to estimation accuracy.
The combinations with $\alpha = 0.0$ yielded especially low RPAs.
This indicates that 
an F0 enhancement spectrogram was effective for vocal F0 estimation.

%
% figure
%

\section{Conclusion}
\label{conclusion}

This paper described a method 
 that performs singing voice separation and vocal F0 estimation
 in a mutually-dependent manner.
%Both tasks can be improved
% by making effective use of their interdependence.
The experimental results showed
 that the proposed method 
 achieves better singing voice separation and vocal F0 estimation
 than conventional methods do.
The singing voice separation of the proposed method
 was also better than that of several state-of-the-art methods in MIREX 2014,
 which is an international competition in music analysis.
In the experiments on vocal F0 estimation,
 the proposed method outperformed two conventional methods
 that are considered to achieve the state-of-the-art performance.
Some parameters of the proposed method significantly
 affect the performances of singing voice separation and vocal F0 estimation,
 and we found that a particular range of those parameters results
 in relatively good performance in various situations.

\revise{
We plan to integrate singing voice separation and vocal F0 estimation in a unified framework.
Since the proposed method performs these tasks in a cascading manner,
 separation and estimation errors are accumulated.
One promising way to solve this problem 
 is to formulate a unified likelihood function to be maximized
 by interpreting the proposed method
 from a viewpoint of probabilistic modeling. 
To discriminate singing voices from musical instrument sounds
 that have sparse and non-repetitive structures in the TF domain like singing voices,
 we attempt to focus on both the structural and timbral characteristics of singing voices
 as in \cite{salamon:2014}. 
It is also important to conduct subjective evaluation to investigate the relationships
 between the conventional measures (SDR, SIR, and SAR) and the perceptual quality.
}%

\section*{Acknowledgment}

The study was supported by
 JST OngaCREST Project, 
 JSPS KAKENHI 24220006, 26700020, and 26280089, 
 and Kayamori Foundation.

\ifCLASSOPTIONcaptionsoff
  \newpage
\fi

\bibliographystyle{./IEEEtran}
\bibliography{taslp2015}


%% \begin{IEEEbiography}{Michael Shell}
%% Biography text here.
%% \end{IEEEbiography}

%% \begin{IEEEbiographynophoto}{Yukara Ikemiya}
%% Biography text here.
%% \end{IEEEbiographynophoto}

%% \begin{IEEEbiographynophoto}{Kazuyoshi Yoshii}
%% Biography text here.
%% \end{IEEEbiographynophoto}

%% \begin{IEEEbiographynophoto}{Katsutoshi Itoyama}
%% Biography text here.
%% \end{IEEEbiographynophoto}


%\vfill


\end{document}